\documentclass[sigconf]{acmart}

\usepackage{algorithmic}
\usepackage{algorithm}
\usepackage{array}
\usepackage{verbatim}
\usepackage{booktabs}       % \toprule \midrule \cmidrule
\usepackage{multirow}       % \multirow
\usepackage{makecell}       % \makecell
\usepackage{threeparttable} % threeparttable environment
\usepackage{subcaption}
\usepackage[table]{xcolor}
\usepackage{pifont}
\usepackage[dvipsnames]{xcolor}
\usepackage{multirow}
\usepackage{balance} 

\definecolor{TableBlue}{RGB}{210, 230, 255}
\newcommand{\cmark}{\textcolor{ForestGreen}{\ding{51}}}
\newcommand{\xmark}{\textcolor{BrickRed}{\ding{55}}}

\AtBeginDocument{%
  }

\copyrightyear{2026}
\acmYear{2026}
\setcopyright{cc}
\setcctype{by}
\acmConference[MM '26]{Proceedings of the 34th ACM International Conference on Multimedia}{November 10--14, 2026}{Rio de Janeiro, Brazil}
\acmBooktitle{Proceedings of the 34th ACM International Conference on Multimedia (MM '26), November 10--14, 2026, Rio de Janeiro, Brazil}
\acmDOI{10.1145/3767308.3836107}
\acmISBN{979-8-4007-2213-4/2026/11}
\begin{document}

\title[PhyAVBench: A Challenging Audio Physics-Sensitivity Benchmark for Physically Grounded T2AV Generation]{PhyAVBench: A Challenging Audio Physics-Sensitivity Benchmark for Physically Grounded Text-to-Audio-Video Generation}

\author{Tianxin Xie}
\authornote{These authors contributed equally to this work.}
\email{txie151@connect.hkust-gz.edu.cn}
\author{Wentao Lei}
\authornotemark[1]
\email{wlei117@connect.hkust-gz.edu.cn}
\author{Kai Jiang}
\authornotemark[1]
\email{kai.jiang@connect.hkust-gz.edu.cn}
\author{Guanjie Huang}
\authornotemark[1]
\email{ghuang565@connect.hkust-gz.edu.cn}
\author{Pengfei Zhang}
\authornotemark[1]
\email{pzhang176@connect.hkust-gz.edu.cn}
\affiliation{%
  \institution{The Hong Kong University of Science and Technology (Guangzhou)}
  \city{Guangzhou}
  \country{China}
}

\author{Chunhui Zhang}
\email{chunhui.zhang@sjtu.edu.cn}
\author{Fengji Ma}
\email{fma421@connect.hkust-gz.edu.cn}
\author{Haoyu He}
\email{hhe152@connect.hkust-gz.edu.cn}
\author{Han Zhang}
\email{hzhang743@connect.hkust-gz.edu.cn}
\author{Jiangshan He}
\email{jhe367@connect.hkust-gz.edu.cn}
\affiliation{%
  \institution{The Hong Kong University of Science and Technology (Guangzhou)}
  \city{Guangzhou}
  \country{China}
}

\author{Jinting Wang}
\email{jwang644@connect.hkust-gz.edu.cn}
\author{Linghan Fang}
\email{linghanfang@outlook.com}
\author{Lufei Gao}
\email{lgao638@connect.hkust-gz.edu.cn}
\author{Orkesh Ablet}
\email{aw063@connect.hkust-gz.edu.cn}
\affiliation{%
  \institution{The Hong Kong University of Science and Technology (Guangzhou)}
  \city{Guangzhou}
  \country{China}
}

\author{Peihua Zhang}
\email{patrickzhang2333@gmail.com}
\affiliation{%
  \institution{Tencent}
  \city{Shenzhen}
  \country{China}
}

\author{Ruolin Hu}
\email{rhu698@connect.hkust-gz.edu.cn}
\author{Shengyu Li}
\email{sli802@connect.hkust-gz.edu.cn}
\affiliation{%
  \institution{The Hong Kong University of Science and Technology (Guangzhou)}
  \city{Guangzhou}
  \country{China}
}

\author{Weilin Lin}
\email{wlin760@connect.hkust-gz.edu.cn}
\author{Xiaoyang Feng}
\email{xfeng733@connect.hkust-gz.edu.cn}
\affiliation{%
  \institution{The Hong Kong University of Science and Technology (Guangzhou)}
  \city{Guangzhou}
  \country{China}
}

\author{Xinyue Yang}
\email{xyang360@connect.hkust-gz.edu.cn}
\author{Yan Rong}
\email{yrong854@connect.hkust-gz.edu.cn}
\author{Yanyun Wang}
\email{yynwang@connect.hku.hk}
\author{Zihang Shao}
\email{zshao199@connect.hkust-gz.edu.cn}
\author{Zelin Zhao}
\email{zzhao450@connect.hkust-gz.edu.cn}
\affiliation{%
  \institution{The Hong Kong University of Science and Technology (Guangzhou)}
  \city{Guangzhou}
  \country{China}
}

\author{Chenxing Li}
\email{chenxingli@tencent.com}
\author{Shan Yang}
\email{shaanyang@tencent.com}
\author{Wenfu Wang}
\email{wangwenfubuaa@163.com}
\author{Meng Yu}
\email{raymondmyu@global.tencent.com}
\author{Dong Yu}
\email{dongyu@ieee.org}
\affiliation{%
  \institution{Tencent}
  \city{Shenzhen}
  \country{China}
}

\author{Li Liu}
\correspondingauthor
\email{avrillliu@hkust-gz.edu.cn}
\affiliation{%
  \institution{The Hong Kong University of Science and Technology (Guangzhou)}
  \city{Guangzhou}
  \country{China}
}

\renewcommand{\shortauthors}{Tianxin Xie et al.}
%% No italics, no superscripts, not anonymous
%% Use footnote or author note to identify equal contribution, shared contribution, and/or contact author info


\begin{abstract}
Text-to-audio-video (T2AV) generation is central to applications such as filmmaking and world modeling. However, current models often fail to produce physically plausible sounds. Previous benchmarks in this area primarily focus on audio-video temporal synchronization, while largely overlooking explicit evaluation of audio-physics grounding, thereby limiting the study of physically plausible audio-visual generation. To address this issue, we present \textbf{PhyAVBench}, the first benchmark designed to systematically evaluate the audio-physics grounding capabilities of T2AV, image-to-audio-video (I2AV), and video-to-audio (V2A) models. PhyAVBench offers \textbf{PhyAV-Sound-11K}, a new dataset of 25.5 hours of 11,605 audible videos collected from 184 participants to ensure diversity and avoid data leakage. It contains 337 paired-prompt groups with controlled physical variations that drive sound differences, each grounded with an average of 17 videos and spanning 6 audio-physics dimensions and 41 fine-grained test points, from basic phenomena (e.g., collision) to complex effects (e.g., Helmholtz resonance). Each video includes step-by-step audio-physics reasoning, and each prompt pair is annotated with the physical factors underlying their acoustic differences. Importantly, unlike prior benchmarks that cannot measure sensitivity to underlying acoustic conditions, PhyAVBench leverages paired text prompts to evaluate this capability. We term this evaluation paradigm the Audio-Physics Sensitivity Test (APST) and introduce a novel metric, the \textbf{Contrastive Physical Response Score (CPRS)}, which quantifies the acoustic consistency between generated videos and their real-world counterparts. We conduct a comprehensive evaluation of 17 state-of-the-art (SOTA) models across T2AV, I2AV, and V2A tasks, along with human studies involving 74 participants, which show a strong positive correlation with the CPRS metric. Our results reveal that even leading commercial models struggle with fundamental audio-physical phenomena, exposing a critical gap beyond audio-visual synchronization and pointing to future research directions. We hope PhyAVBench will serve as a foundation for advancing physically grounded audio-visual generation. Code and datasets are available at \url{https://github.com/imxtx/PhyAVBench}.
\end{abstract}



% \keywords{Audio-video generation, T2AV, benchmark, physics-sensitivity, physically grounded}



\maketitle

\section{Introduction}

Text-to-audio-video (T2AV) generation has attracted significant attention since the release of Sora 2~\cite{sora2}, which inspired a wide range of applications such as filmmaking, advertising, and world modeling.
Before this milestone, several approaches had already explored simultaneous T2AV, e.g., SyncFlow~\cite{liu2024syncflow}, JavisDiT~\cite{liu2025javisdit}, UniVerse-1~\cite{wang2025universe}, and Ovi~\cite{low2025ovi}. 
Compared with video-only~\cite{singer2022make,wu2023tune,hong2022cogvideo,wang2023modelscope,wu2024motionbooth}, or video-to-audio generation methods~\cite{wang2024frieren,wang2024tiva,ren2025sta,wang2025vaflow,cheng2025lova}, these models have substantially advanced T2AV by jointly synthesizing visual and auditory content, thereby making the creation of audio-visual media more accessible to end users.
However, existing methods still struggle to generate physically plausible audio, often producing unrealistic sounds or exhibiting audio-visual asynchrony~\cite{wang2024av,kim2024versatile,xing2024seeing,zhao2025uniform,haji2025av,ishii2025simple}.
At the same time, the field lacks clear methodological guidance: researchers face difficulties in designing and training audio-physics-aware models, and current benchmarks provide limited insight into the specific deficiencies of existing systems in terms of audio-physical grounding.
These limitations collectively hinder further progress and motivate the need to address the following issues of existing benchmarks and datasets.

\begin{figure}[!t]
  \centering
  \includegraphics[width=\linewidth]{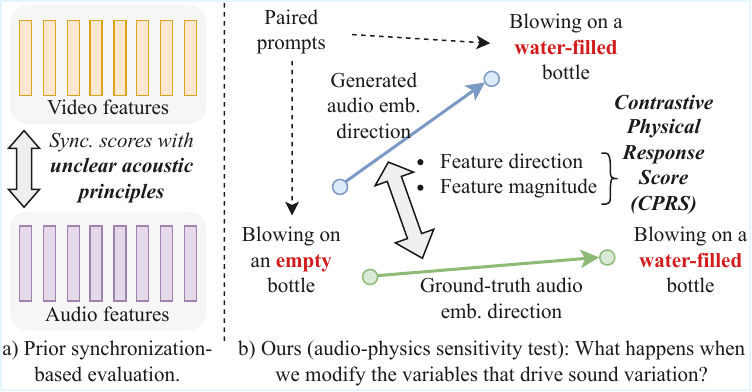}
  \caption{The proposed audio-physics sensitivity test.}
  \label{fig:cprs}
\end{figure}

\begin{table*}[!t]
% \fontsize{8.5pt}{9.5pt}\selectfont
% \setlength{\tabcolsep}{2pt}
% \renewcommand{\arraystretch}{0.8}
\centering
\caption{Comparison with the existing audio-video generation benchmarks.}
\resizebox{1.0\textwidth}{!}{
\begin{threeparttable}
    \begin{tabular}{lccccccccccc}
    \toprule
    & \multicolumn{6}{c}{Acoustic Physical Dimensions} & \multirow{2}{*}[-3pt]{\makecell{Controlled\\ Setting}} & \multirow{2}{*}[-3pt]{\makecell{Newly\\ Collected}} & \multirow{2}{*}[-3pt]{\makecell{\#GT Videos\\ per Prompt}} & \multirow{2}{*}[-3pt]{\makecell{Evaluation\\ Metric}} \\
    \cmidrule(rl){2-7}
    & \makecell{Sound Source\\ Mechanism} & \makecell{Fluid \&\\ Aerodynamics} & \makecell{Sound\\ Propagation} & \makecell{Observer\\ Physics} & \makecell{Time \&\\ Causality} & \makecell{Complex\\ Coupling} & & & & \\
    \midrule
    TAVGBench~\cite{mao2024tavgbench} & \xmark & \xmark & \xmark & \xmark & \xmark & \xmark & \xmark & \xmark & 1 & AV-Align\\
    SAVGBench~\cite{shimada2024savgbench} & \xmark & \xmark & \xmark & \xmark & \cmark & \xmark & \xmark & \xmark & - & AV \& Spatial-Align \\ 
    Verse-Bench~\cite{wang2025universe} & \xmark & \xmark & \xmark & \xmark & \xmark & \xmark & \xmark & \cmark & 1 & AV-Align\\
    JavisBench~\cite{liu2025javisdit} & \xmark & \xmark & \xmark & \xmark & \xmark & \xmark & \xmark & \cmark (partial) & 1 & AV-Align \\
    VABench~\cite{hua2025vabench} & \cmark & \cmark & \xmark & \xmark & \cmark & \xmark & \xmark & \xmark & 0 & AV \& Stereo Align \\
   T2AV-Compass~\cite{cao2025t2av} & \cmark & \xmark & \xmark & \xmark & \cmark & \xmark & \xmark & \xmark & 1 & \makecell{AV-Align} \\
    \midrule
    \textbf{PhyAVBench (Ours)} & \cmark & \cmark & \cmark & \cmark & \cmark & \cmark & \cmark & \cmark & 2$\sim$40 & \makecell{AV-Align \& Physics Sensitivity}\\ 
    \bottomrule
\end{tabular}
\end{threeparttable}}
\label{tab:comparison}
\end{table*}

\textit{i) Limited acoustics coverage.}
As shown in Tab.~\ref{tab:comparison}, Current benchmarks predominantly emphasize high-level audio-visual alignment, such as synchronization and semantic consistency, while covering only a narrow subset of acoustic phenomena \cite{cao2025t2av,shimada2024savgbench,hua2025vabench}.
Crucial aspects of real-world sound generation, such as material-dependent timbre, resonance, force-induced amplitude variation, sound propagation, and environmental effects, are largely underexplored.
As a result, existing evaluations fail to comprehensively test whether models can generalize across diverse acoustic conditions or capture the underlying physical principles that govern sound generation.

\textit{ii) Reusing existing datasets.}
Many benchmarks \cite{mao2024tavgbench,shimada2024savgbench,hua2025vabench,cao2025t2av} are constructed by repurposing existing video or audio-visual datasets (e.g., AudioSet~\cite{gemmeke2017audio}) that were not originally designed for studying audio-physical relationships.
This practice introduces several limitations, including uncontrolled variables, inconsistent recording conditions, and potential data leakage from training corpora.
More importantly, such datasets lack carefully designed counterfactual or controlled variations, making it difficult to isolate specific physical factors and rigorously assess models' sensitivity to them.

\textit{iii) Lack of fine-grained acoustics reasoning annotation.}
Existing datasets \cite{gemmeke2017audio,primus2025tacos,xie2025audiotime,wei2022desed} typically provide coarse-grained labels (e.g., event categories or captions) without explicitly annotating the underlying acoustic factors or causal relationships that produce sound.
This absence of fine-grained annotations, such as material type, interaction force, spatial configuration, or temporal causality, limits both the interpretability of model behavior and the ability to conduct diagnostic evaluation. Consequently, it remains unclear whether models truly capture audio-physical relationships or merely rely on superficial correlations.

\textit{iv) Lack of a specialized audio-grounding evaluation protocol.}
Current evaluation protocols \cite{shimada2024savgbench,hua2025vabench,liu2025javisdit,wang2025universe,cao2025t2av,mao2024tavgbench} largely rely on global perceptual metrics or human judgments, which are insufficient for assessing audio-physical grounding.
As shown in Fig.~\ref{fig:cprs}, they do not explicitly test whether models respond correctly to controlled changes in physical conditions, nor do they quantify the direction and magnitude of such responses.
Without a dedicated evaluation paradigm that probes models' sensitivity to physically meaningful variations, it is difficult to systematically diagnose their shortcomings or guide the development of physics-aware generative models.

To address these gaps and advance future research, we present \textbf{PhyAVBench}, a new challenging benchmark for physically ground-ed T2AV generation that evaluates models' understanding of audio-related physical laws by measuring their sensitivity to variations in acoustic conditions across a wide range of scenes.
As shown in Fig.~\ref{fig:cprs}, PhyAVBench comprises groups of paired T2AV prompts, where the two prompts in each group differ only in a single controlled variable that induces corresponding variations in sound.
For example, a paired prompt contrasts knocking on wood with knocking on metal under identical visual conditions.
The induced directional change in timbre and resonance directly reflects the underlying material-dependent acoustic properties, which is precisely what our physics-sensitivity test is designed to capture.
We term this evaluation paradigm the \textbf{Audio-Physics Sensitivity Test (APST)} and compute a novel metric, i.e., \textbf{Contrastive Physical Response Score (CPRS)}, for each prompt group, enabling quantitative assessment of both directional alignment and magnitude consistency in the physical transitions.
This design ensures that models cannot rely on superficial visual cues or semantic priors alone, but must instead capture the underlying physical relationships between the incentives (e.g., actions and materials) and the resulting sounds in order to generate physically plausible audio.

To support comprehensive evaluation, we present PhyAV-Sound-11K, a high-fidelity 25.5-hour dataset of 11,605 newly recorded videos captured in controlled environments, explicitly designed to prevent data leakage.
This dataset features diverse participants (184 in total), recording scenarios, fine-grained human-refined annotations, and coverage of diverse acoustic phenomena.
For each prompt pair, multiple ground-truth videos are recorded under strictly controlled acoustic conditions, enabling reliable comparison across subtle physical variations.
All prompts and videos undergo thorough review and validation by multiple human experts to guarantee both quality and cross-sample consistency.

We extensively evaluate over 17 SOTA models (e.g., Sora 2 and Veo 3.1) and observe a pronounced performance gap.
We find that even leading commercial systems struggle to capture fine-grained physical transitions; for example, Sora 2 achieves a CPRS of only 0.4512, highlighting the urgent need for physics-aware generative modeling.
Our contributions are summarized as follows:

1) We introduce PhyAVBench, the first benchmark for evaluating audio-physics grounding in T2AV, I2AV, and V2A models, featuring a hierarchy of 6 dimensions and 41 fine-grained acoustic test cases across diverse real-world scenarios.

2) We propose the APST founded on a controlled-variable paradigm. By utilizing human-refined paired prompts, this approach isolates physical causality from semantic priors, enabling a precise assessment of a model's sensitivity to underlying acoustic laws.

3) We propose a novel CPRS to measure the alignment of generated audio features with ground-truth physical trends. Comprehensive evaluation and analysis demonstrate that CPRS effectively captures the subjective dimension of physical rationality, showing a strong positive correlation with expert human perception.

4) We release PhyAV-Sound-11K, a curated dataset of 11,605 original audio-video pairs. Unlike benchmarks relying on reused data, all our samples are newly recorded in controlled settings, with each prompt supported by an average of 17 independent ground-truth samples to ensure statistical reliability and zero data leakage.

\section{Related Work}
\label{sec:related}

\subsection{Text-to-Audio-Video Models}

\noindent\textbf{Unified T2AV Models.}
Text-to-audio-video models jointly synthesize coherent audiovisual content from text, typically using diffusion or autoregressive architectures to align text, audio, and video.
Commercial systems such as Sora 2~\cite{sora2}, Veo 3.1~\cite{veo3.1}, Seedance 1.5 Pro~\cite{seedance1.5pro}, Kling v2.6~\cite{Kling-V2-6}, and Wan 2.6~\cite{Wan2.6} mark a shift from cascaded pipelines to unified audio-visual generation. Large-scale multimodal training and cross-modal fusion enable them to generate coherent, high-fidelity content in open-domain scenarios.
Academic research explores unified architectures for audio-visual alignment. Seeing and Hearing~\cite{xing2024seeing}, CMC-PE~\cite{ishii2025simple}, and Kim et al.~\cite{kim2024versatile} study latent alignment, timestep control, and mixture-of-noise training. SyncFlow~\cite{liu2024syncflow}, AV-Link~\cite{haji2025av}, JavisDiT~\cite{liu2025javisdit}, and AV-DiT~\cite{wang2024av} improve cross-modal synchronization through flow matching, temporal priors, and modality-specific adapters. UniVerse-1~\cite{wang2025universe} stitches pretrained modality experts, whereas UniForm~\cite{zhao2025uniform} operates in a shared multimodal latent space.

\noindent\textbf{Video-to-Audio Models.}
Video-to-audio models target semantically relevant and temporally synchronized Foley generation. FoleyCrafter~\cite{zhang2024foleycrafter}, HunyuanVideo-Foley~\cite{shan2025hunyuanvideo}, and MMAudio~\cite{cheng2025mmaudio} employ semantic conditioning, representation alignment, and flow matching, respectively, while ThinkSound~\cite{liu2025thinksound} introduces reasoning for improved logical consistency. However, modeling physical causality between visual interactions and sounds remains challenging.

\subsection{Text-to-Audio-Video Benchmarks}

\noindent \textbf{Unified Cross-Modal Evaluation.} 
To facilitate the advancement of T2AV systems, the evaluation paradigm has shifted from unimodal quality assessment to holistic benchmarks that scrutinize the semantic and temporal alignment between visual and auditory modalities. Initial efforts, such as TAVGBench~\cite{mao2024tavgbench} and Verse-Bench~\cite{wang2025universe}, laid the groundwork by emphasizing the synchronization and semantic consistency of synthesized content under complex prompts, while JavisBench~\cite{liu2025javisdit} highlighted the challenges of maintaining cross-modal coherence in open-ended joint generation tasks. Most recently, frameworks like VABench~\cite{hua2025vabench} have further unified these protocols, establishing comprehensive standards that integrate objective signal-level fidelity with subjective instruction-following metrics to provide a multi-dimensional assessment. 

\noindent \textbf{Task-Specific Evaluation.} 
Beyond general generation quality, significant research efforts have been directed towards specific dimensions of audio-visual correspondence. Addressing temporal precision, PEAVS~\cite{goncalves2024peavs} establishes a perceptual metric grounded in human opinion, while Mo et al.~\cite{mo2024text} propose quantitative standards for visual alignment and temporal consistency. In the spatial domain, SAVGBench~\cite{shimada2024savgbench} and Real Acoustic Fields~\cite{chen2024real} evaluate the alignment of spatial audio with visual sources and room acoustics, respectively. Furthermore, specialized benchmarks target distinct generative sub-tasks: FoleyBench~\cite{dixit2025foleybench} focuses on category diversity and causality in video-to-audio synthesis; DualDub~\cite{tian2025dualdub} addresses the joint synthesis of speech and background tracks; and LVAS-Bench~\cite{zhang2025long} tackles the consistency challenges inherent in long-form video-to-audio synthesis. However, these benchmarks provide limited coverage of explicit audio-physical principles and lack controlled-variable settings, making them insufficient for diagnosing physical understanding. We therefore introduce PhyAVBench to systematically evaluate audio-physics grounding in T2AV models, with comparisons provided in Tab.~\ref{tab:comparison}.

\section{PhyAVBench}
\label{sec:bench}

\subsection{Overview}

PhyAVBench is designed to systematically evaluate the audio-physics grounding capability of T2AV models.
It comprises 25.5 hours of 11,605 newly recorded videos, captured by 184 participants and organized into 337 groups of human-refined prompts.
The benchmark features controlled physical factors, paired comparisons, and diverse real-world recordings, enabling fine-grained assessment of models' physical consistency beyond surface-level realism.

\begin{figure*}[t]
    \centering
    \includegraphics[width=\linewidth]{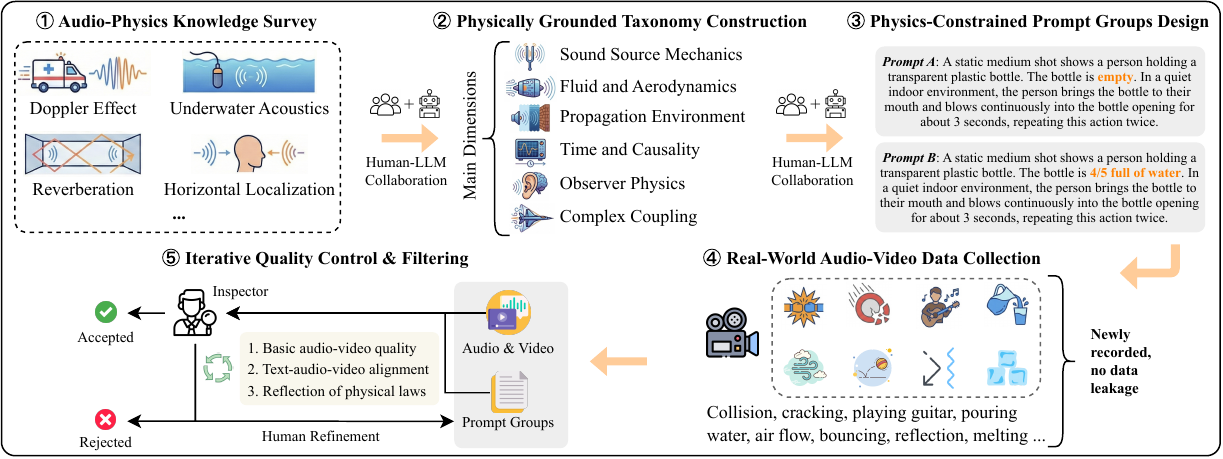}
    \caption{The data curation pipeline of PhyAVBench.}
    \label{fig:pipeline}
\end{figure*}

\noindent\textbf{Audio Physics Coverage.}
PhyAVBench spans 6 major audio-physics dimensions and contains 41 fine-grained test points.
These test points range from common acoustic phenomena, such as reverberation and collision sound, to more complex effects, including vortex shedding and Helmholtz resonance, forming a comprehensive evaluation matrix.
Detailed descriptions and examples are provided in the \textbf{supplementary material}.

\noindent\textbf{Annotations.}
PhyAVBench provides expert-verified, multi-level annotations of the causal sound-generation process, covering interacting objects, materials, action dynamics, environmental conditions, and acoustic outcomes. Each paired prompt further identifies the controlled variable, the resulting acoustic change, and its physical cause. By capturing both \emph{what} sound is produced and \emph{why}, these annotations support diagnostic evaluation of models' audio-physics grounding and sensitivity to controlled physical variations.

\subsection{Data Curation Pipeline}

As shown in Fig.~\ref{fig:pipeline}, PhyAVBench is constructed through a five-stage pipeline integrating audio-physics research, taxonomy design, prompt construction, video collection, and iterative quality control:
\textbf{1) Audio-Physics Knowledge Survey.}
We systematically survey the physics and acoustics of sound generation and propagation, covering key principles, mechanisms, and observable audio--visual scenarios. LLM-assisted brainstorming provides broad candidate coverage, while human experts remove infeasible, redundant, or irrelevant cases and consolidate related concepts. The resulting knowledge base provides a principled foundation for the taxonomy and benchmark.
\textbf{2) Physically Grounded Taxonomy Construction.}
Based on this knowledge, we construct a hierarchical taxonomy spanning \textit{Sound Source Mechanics}, \textit{Fluid and Aerodynamics}, \textit{Sound Propagation Environment}, \textit{Observer Physics}, \textit{Time and Causality}, and \textit{Complex Coupling}, each decomposed into fine-grained, controllable test points. LLMs propose the initial structure, which experts refine to resolve ambiguities, remove redundancies, and ensure physical interpretability. This collaboration produces a comprehensive, non-overlapping benchmark hierarchy.
\textbf{3) Physics-Constrained Prompt Group Design.}
Guided by the taxonomy, LLMs generate prompts describing physical scenarios with controlled variables such as force, speed, contact area, material stiffness, and fluid flow rate. The prompts are refined to eliminate linguistic ambiguity, subjective descriptions, and explicit acoustic outcomes. Experts further ensure that each prompt pair differs in only one physical factor while all other conditions remain fixed, enabling precise sensitivity analysis of T2AV models.
\textbf{4) Real-World Audio-Video Data Collection.}
We record new videos for every prompt, avoiding the data leakage risks of prior benchmarks~\cite{liu2025javisdit,mao2024tavgbench,hua2025vabench,shimada2024savgbench} that reuse datasets such as AudioSet~\cite{gemmeke2017audio}. Recordings isolate the target variable under controlled conditions while maintaining diversity across instances, performers, and devices. Multiple independent samples per prompt improve statistical reliability and reduce the effects of noise and recording artifacts.
\textbf{5) Iterative Quality Control and Filtering.}
LLMs first screen prompts and videos for ambiguity, unintended confounders, and control-variable violations. Human experts then verify physical consistency, text--audio--video alignment, and real-world plausibility. Problematic samples are removed, revised, or re-collected until they satisfy strict quality standards, ensuring reliable physically grounded T2AV evaluation.

\section{Evaluation}
\label{sec:eval}

\subsection{Experiment Settings}
PhyAVBench is specifically designed to evaluate the audio-physical grounding capability of audio-video generation models.
In current settings, two primary paradigms are considered, i.e., T2AV and I2AV generation, both of which generate audio jointly with the visual content conditioned on text prompts.
In addition, PhyAVBench applies to assessing the audio-physical grounding ability of V2A models.
We do not consider traditional text-to-voiceless video or audio-to-video generation tasks. 
We evaluate both SOTA commercial and open-source T2AV~\cite{sora2,veo3.1,seedance1.5pro,Kling-V2-6,Wan2.6,liu2026javisdit++,liu2025javisdit,low2025ovi,hacohen2026ltx}, I2AV~\cite{wang2025universe,team2026mova,low2025ovi,hacohen2026ltx}, and representative V2A~\cite{zhang2024foleycrafter,shan2025hunyuanvideo,cheng2025mmaudio,liu2025thinksound} models. 

For closed-source models, we evaluate only the T2AV task due to limited budget, randomly sampling 100 prompt pairs from the dataset.
For open-source models, we consider all 337 prompt groups.
For the I2AV setting, the first frame of the ground-truth video is used as the input image. For the V2A task, we provide the ground-truth videos with audio removed as input to the models.
For CPRS, FVD, and KVD, we use the complete set of ground-truth videos as references during evaluation.

\begin{table*}[t] 
% \fontsize{8.5pt}{10pt}\selectfont
% \setlength{\tabcolsep}{3pt}
\renewcommand{\arraystretch}{0.8}
\centering
\caption{Quantitative evaluation on PhyAVBench. We categorize models by task (T2AV, I2AV, V2A) and source type. Bold and underlined values denote the best and second-best results within each task category, respectively.}
\label{tab:results}
\resizebox{1.0\textwidth}{!}{
\begin{tabular}{l|ccc|cccc|cccc|c}
\toprule
\multirow{2}{*}{Model} & \multicolumn{3}{c|}{AV Quality} & \multicolumn{4}{c|}{Semantic Consistency} & \multicolumn{4}{c|}{AV Synchronization} & Physical \\
\cmidrule(lr){2-4} \cmidrule(lr){5-8} \cmidrule(lr){9-12} \cmidrule(lr){13-13}
 & FVD $\downarrow$ & KVD $\downarrow$ & FAD $\downarrow$ & TV-IB $\uparrow$ & TA-IB $\uparrow$ & CLIP $\uparrow$ & CLAP $\uparrow$ & AV-IB $\uparrow$ & DeSync $\downarrow$ & AVH $\uparrow$ & Javis $\uparrow$ & CPRS $\uparrow$ \\
\midrule
\multicolumn{13}{l}{\textbf{Text-to-Audio-Video (T2AV)}} \\
\midrule
\rowcolor[gray]{0.95} \textit{Commercial (Closed-source)} & & & & & & & & & & & & \\
Sora 2 \cite{sora2} & \cellcolor{TableBlue}\textbf{578.70} & \cellcolor{TableBlue}\textbf{72.04} & 4.25 & \underline{0.3415} & 0.2028 & 0.2713 & 0.3592 & 0.2137 & 0.0370 & 0.2118 & 0.1803 & \cellcolor{TableBlue}\textbf{0.4512} \\
Wan 2.6 \cite{Wan2.6} & \underline{656.62} & \underline{108.25} & \cellcolor{TableBlue}\textbf{1.48} & \cellcolor{TableBlue}\textbf{0.3434} & 0.1820 & \cellcolor{TableBlue}\textbf{0.2792} & \underline{0.3858} & 0.2143 & 0.2645 & 0.2141 & 0.1846 & \underline{0.4120} \\
Seedance 1.5 Pro \cite{seedance1.5pro} & 968.03 & 315.38 & 3.09 & 0.3274 & \cellcolor{TableBlue}\textbf{0.2258} & 0.2663 & 0.3600 & \cellcolor{TableBlue}\textbf{0.2794} & \underline{0.0240} & \underline{0.2754} & \cellcolor{TableBlue}\textbf{0.2412} & 0.4044 \\
Kling v2.6 \cite{Kling-V2-6} & 737.52 & 162.64 & \underline{1.48} & 0.3373 & \underline{0.2172} & \underline{0.2725} & \cellcolor{TableBlue}\textbf{0.4252} & \underline{0.2654} & 0.2070 & \cellcolor{TableBlue}\textbf{0.2756} & \underline{0.2355} & 0.3718 \\
Veo 3.1 \cite{veo3.1} & 755.51 & 189.83 & 3.48 & 0.3394 & 0.1886 & 0.2723 & 0.3393 & 0.2136 & \cellcolor{TableBlue}\textbf{0.0160} & 0.2211 & 0.1950 & 0.3563 \\
\midrule
\rowcolor[gray]{0.95} \textit{Open-source} & & & & & & & & & & & & \\
Ovi \cite{low2025ovi} & \underline{529.44} & 238.87 & \cellcolor{TableBlue}\textbf{2.70} & 0.3113 & 0.0930 & \underline{0.2625} & 0.2635 & \underline{0.1509} & 0.3472 & \underline{0.1578} & \underline{0.1355} & \cellcolor{TableBlue}\textbf{0.3290} \\
LTX \cite{hacohen2026ltx} & 682.38 & \underline{175.86} & 5.76 & 0.3127 & 0.1083 & 0.2619 & 0.2583 & 0.1178 & \cellcolor{TableBlue}\textbf{0.2531} & 0.1332 & 0.1039 & \underline{0.3235} \\
JavisDiT \cite{liu2025javisdit} & 656.82 & 244.63 & \underline{3.29} & \underline{0.3156} & \underline{0.1350} & 0.2610 & \underline{0.2906} & 0.1293 & \underline{0.2893} & 0.1320 & 0.1148 & 0.2938 \\
JavisDiT++ \cite{liu2026javisdit++} & \cellcolor{TableBlue}\textbf{474.62} & \cellcolor{TableBlue}\textbf{76.89} & 4.55 & \cellcolor{TableBlue}\textbf{0.3296} & \cellcolor{TableBlue}\textbf{0.1506} & \cellcolor{TableBlue}\textbf{0.2689} & \cellcolor{TableBlue}\textbf{0.3031} & \cellcolor{TableBlue}\textbf{0.1553} & 0.5427 & \cellcolor{TableBlue}\textbf{0.1776} & \cellcolor{TableBlue}\textbf{0.1498} & 0.2844 \\
\midrule
\multicolumn{13}{l}{\textbf{Image-to-Audio-Video (I2AV)}} \\
\midrule
MOVA \cite{team2026mova} & \underline{157.43} & \underline{14.17} & \cellcolor{TableBlue}\textbf{1.37} & \cellcolor{TableBlue}\textbf{0.3116} & \cellcolor{TableBlue}\textbf{0.1859} & \cellcolor{TableBlue}\textbf{0.2676} & \cellcolor{TableBlue}\textbf{0.3404} & \cellcolor{TableBlue}\textbf{0.2116} & 0.7799 & \cellcolor{TableBlue}\textbf{0.2110} & \cellcolor{TableBlue}\textbf{0.1679} & \cellcolor{TableBlue}\textbf{0.3613} \\
LTX \cite{hacohen2026ltx} & 254.35 & 45.99 & 7.53 & \underline{0.3100} & \underline{0.1077} & 0.2644 & 0.1911 & 0.0964 & \cellcolor{TableBlue}\textbf{0.2984} & 0.1010 & 0.0772 & \underline{0.3285} \\
Ovi \cite{low2025ovi} & \cellcolor{TableBlue}\textbf{122.79} & \cellcolor{TableBlue}\textbf{7.46} & \underline{3.15} & 0.3049 & 0.0911 & \underline{0.2645} & \underline{0.2617} & \underline{0.1526} & \underline{0.3410} & \underline{0.1579} & \underline{0.1373} & 0.3198 \\
UniVerse-1 \cite{wang2025universe} & 340.20 & 32.14 & 5.84 & 0.2573 & 0.0834 & 0.2338 & 0.1575 & 0.0792 & 0.7269 & 0.0806 & 0.0565 & 0.2729 \\
\midrule
\multicolumn{13}{l}{\textbf{Video-to-Audio (V2A)}} \\
\midrule
MMAudio \cite{cheng2025mmaudio} & - & - & \cellcolor{TableBlue}\textbf{1.36} & 0.2945 & \cellcolor{TableBlue}\textbf{0.2330} & 0.2632 & \cellcolor{TableBlue}\textbf{0.3890} & \underline{0.2684} & \underline{0.3165} & \underline{0.2697} & \underline{0.2200} & \cellcolor{TableBlue}\textbf{0.4003} \\
HunyuanVideo-Foley \cite{shan2025hunyuanvideo} & - & - & 1.64 & 0.2950 & 0.2164 & 0.2636 & \underline{0.3695} & \cellcolor{TableBlue}\textbf{0.2721} & \cellcolor{TableBlue}\textbf{0.3072} & \cellcolor{TableBlue}\textbf{0.2727} & \cellcolor{TableBlue}\textbf{0.2260} & \underline{0.3896} \\
ThinkSound \cite{liu2025thinksound} & - & - & \underline{1.60} & 0.2950 & 0.1582 & 0.2634 & 0.3412 & 0.1874 & 0.9284 & 0.1941 & 0.1458 & 0.3186 \\
FoleyCrafter \cite{zhang2024foleycrafter} & - & - & 1.88 & 0.2945 & \underline{0.2204} & 0.2627 & 0.3185 & 0.2520 & 0.9016 & 0.2580 & 0.2101 & 0.3019 \\
\bottomrule
\end{tabular}
}
\end{table*}

\begin{figure*}[t] % 使用 figure* 使图片跨双栏显示
  \centering
  \includegraphics[width=0.9\linewidth]{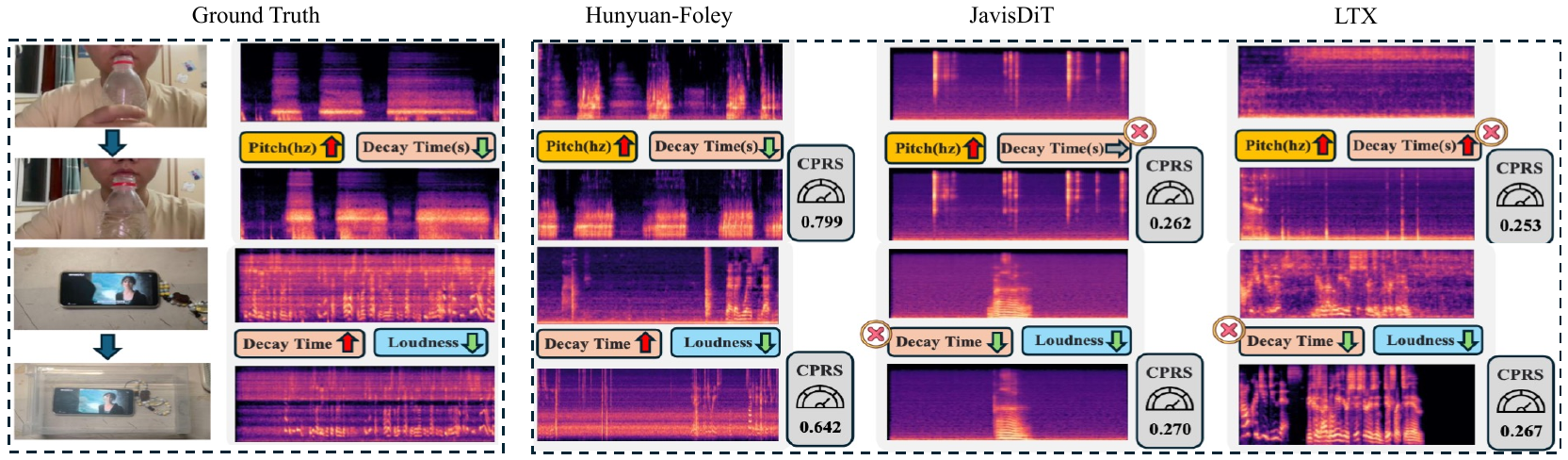} 
  \caption{Qualitative comparison of generated audio-visual results on PhyAVBench. }
  \label{fig:qualitative}
\end{figure*}

\begin{figure*}[t]
    \centering
    \includegraphics[width=0.95\textwidth]{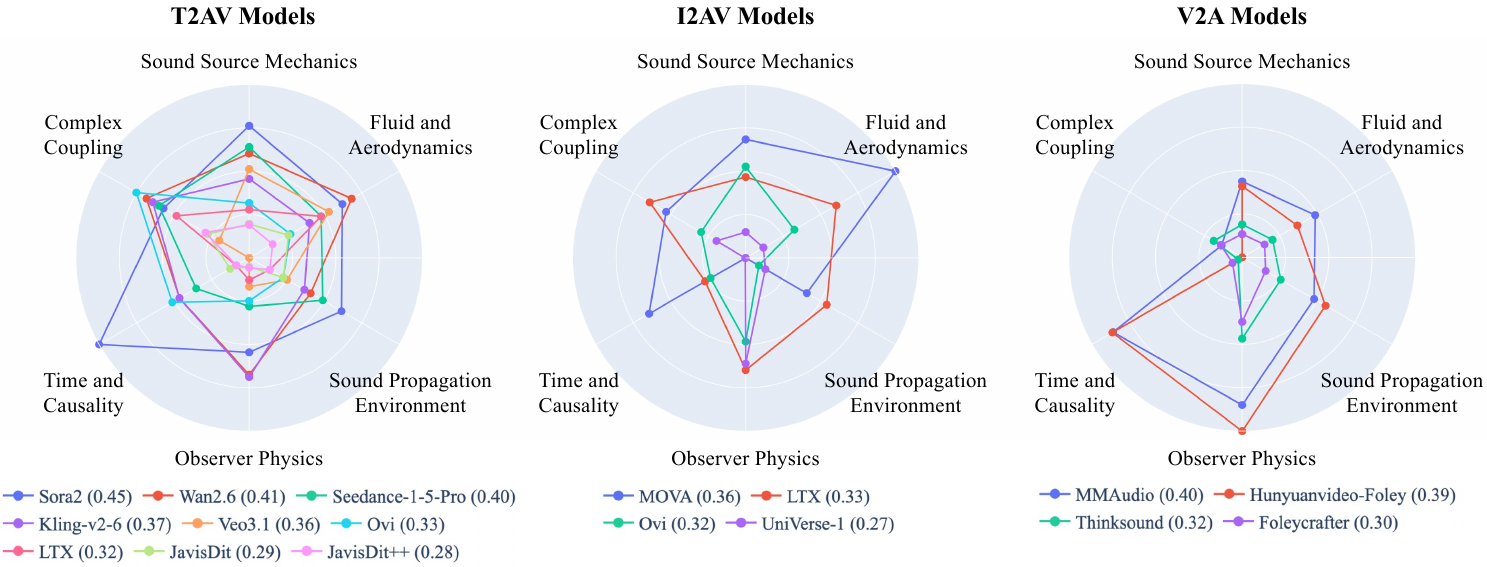}
    \caption{The audio-physics grounding performance of SOTA T2AV, I2AV, and V2A models measured by CPRS.}
    \label{fig:rador}
\end{figure*}

\subsection{Metrics}

\noindent\textbf{Audio-Physics Sensitivity Test (APST).}
The APST evaluates whether models respond correctly to physically meaningful changes in text prompts, focusing on directional consistency and magnitude alignment rather than absolute audio quality.
It uses paired or grouped prompts that vary only one audio-related physical variable (e.g., size, material, airflow), where the expected sound change is physically well-defined.
A model is considered physically sensitive if the direction and magnitude of its generated audio align with real-world, physics-grounded trends.
By comparing these two factors with those from ground-truth videos, APST isolates and assesses a model's audio-physical understanding independent of high-level audio-video fidelity or synchronization. 

To this end, we propose \textbf{Contrastive Physical Response Score (CPRS)}, a metric that evaluates the audio-physics grounding ability of T2AV generation by jointly measuring directional alignment and magnitude consistency of physical feature transitions in a shared embedding space.
We formalize the evaluation as follows:

\textbf{1) Compute Ground-Truth Physical Direction.}
Let $\Phi(\cdot)$ denote the audio encoder of a pretrained CLAP model~\cite{wu2023large}, which maps an audio sample into a feature embedding.
For a pair of prompts $P_a$ and $P_b$ corresponding to different physical conditions, we compute the mean embeddings $(\mathbf{e}_{a, \text{\text{gt}}}, \mathbf{e}_{b, \text{\text{gt}}})$ of $N$ pairs of ground-truth audio samples $(a_{\text{gt}, i}, a_{\text{gt}, i})$:
\begin{equation}
    \mathbf{e}_{a, \text{\text{gt}}} = \frac{1}{N} \sum_{i=1}^{N} \Phi(a_{\text{gt}, i}), \quad 
    \mathbf{e}_{b, \text{\text{gt}}} = \frac{1}{N} \sum_{i=1}^{N} \Phi(b_{\text{gt}, i}).
\end{equation}
Then, we define the ground-truth physical direction $\mathbf{v}_{\text{\text{gt}}}$ as follows:
\begin{equation}
    \mathbf{v}_{\text{\text{gt}}} = \mathbf{e}_{b, \text{\text{gt}}} - \mathbf{e}_{a, \text{\text{gt}}}.
\end{equation}

\textbf{2) Compute Generated Direction.}
Given paired audio samples $a_{\text{gen}}$ and $b_{\text{gen}}$ extracted from the generated video pair, we extract their embeddings using the same encoder and compute the generated embedding direction $\mathbf{v}_{\text{gen}}$ as follows:
\begin{equation}
    \mathbf{v}_{\text{gen}} = \Phi(b_{\text{gen}}) - \Phi(a_{\text{gen}}).
\end{equation}

\textbf{3) CPRS Computation.}
We first measure the directional alignment between the generated and ground-truth audio samples using cosine similarity $c$ and normalize it to $[0,1]$:
\begin{equation}
    c = \frac{1}{2} \left( 
    \frac{\mathbf{v}_{\text{gen}} \cdot \mathbf{v}_{\text{\text{gt}}}}{\|\mathbf{v}_{\text{gen}}\| \|\mathbf{v}_{\text{\text{gt}}}\|} + 1 
    \right) \in [0,1].
\end{equation}
To further evaluate whether the magnitude of the generated physical change matches the ground truth, we compute the projection of $\mathbf{v}_{\text{gen}}$ onto $\mathbf{v}_{\text{\text{gt}}}$ and normalize it by the squared norm of $\mathbf{v}_{\text{\text{gt}}}$:
\begin{equation}
    p = \frac{\mathbf{v}_{\text{gen}} \cdot \mathbf{v}_{\text{\text{gt}}}}{\|\mathbf{v}_{\text{\text{gt}}}\|^2}.
\end{equation}
This quantity represents the scaling factor of $\mathbf{v}_{\text{gen}}$ along the ground-truth direction, where $p=1$ indicates perfect magnitude alignment.
We then apply a Gaussian kernel centered at $1$:
\begin{equation}
    f(p) = \exp\left(-k (p - 1)^2 \right) \in [0, 1],
\end{equation}
which assigns the highest score when the projected magnitude matches the ground truth and smoothly penalizes positive or negative deviations.
We set the scaling factor $k = 5$ to ensure a smooth transition of values.
Finally, the CPRS score is defined as:
\begin{equation}
    \text{CPRS} = \frac{1}{2} \left( c + f(p) \right) \in [0,1].
\end{equation}

\noindent\textbf{Interpretation.}
The cosine term $c$ evaluates whether the generated feature transition follows the correct \emph{direction} of physical change, while the Gaussian-normalized projection term $f(p)$ measures whether the \emph{magnitude of change along that direction} matches the ground truth. By combining both terms, CPRS favors generated samples that are both directionally aligned and physically consistent in response strength.

\noindent\textbf{Audio-Video Quality Test.}
Following ~\cite{liu2025javisdit, mao2024tavgbench}, we report \textbf{FVD}, \textbf{KVD}, and \textbf{FAD} to evaluate generation quality. FVD and KVD measure the visual and temporal consistency and FAD assesses audio.

\noindent\textbf{Audio-Video Synchronization Test.}
To evaluate the temporal synchrony and cross-modal alignment of the generated audio and videos, we follow the evaluation metrics in JavisDiT~\cite{liu2025javisdit}. We employ \textbf{JavisScore} to measure spatiotemporal synchronization in diverse, real-world contexts. In addition to JavisScore, we evaluate global cross-modal consistency using \textbf{AV-IB} (Audio-Visual ImageBind similarity). We employ \textbf{DeSync} to estimate the degree of temporal shift and \textbf{AVH} (Audio-Visual Hits) to measure the anchoring accuracy of acoustic events. Together, these metrics measure whether generated sounds are synchronized with visual dynamics. 

\noindent\textbf{T2AV Semantic Alignment and Quality Assessment.}
To evaluate the semantic consistency and conceptual fidelity of the generated audio-visual content relative to the text prompts, we utilize existing modality-specific cross-modal pre-trained models to assess their alignment. For text-audio semantic alignment, following~\cite{huang2023make}, we use CLAP\footnote{\url{https://huggingface.co/laion/clap-htsat-unfused}}: generated audio and the original text prompt are encoded with the same checkpoint, and their normalized embeddings are compared by cosine similarity. TA-IB likewise compares the original prompt text with generated-audio ImageBind embeddings.
Similarly, for the text-video semantic alignment, following~\cite{wu2021godiva}, we leverage the CLIPSIM (average CLIP~\cite{radford2021learningtransferablevisualmodels} similarity between video frames and text) to measure the quality of text-video alignment.
Additionally, following prior works~\cite{liu2025javisdit,mao2024tavgbench}, we report FVD, KVD, and FAD to measure audio and video generation quality.

\noindent\textbf{Human Evaluation.}
To further validate the audio-physical plausibility of the generated content, we conducted a human evaluation focusing on \textit{text consistency}, \textit{audio-visual synchronization}, and \textit{physical variation response}, reporting text consistency mean-opinion-score (\textbf{TC-MOS}), \textbf{AV-Sync-MOS}, and \textbf{PVR-MOS}, respectively. All metrics are rated on a 1--5 Likert scale, where higher scores indicate better performance.
TC-MOS measures how well the generated audio-visual content aligns with the textual prompt.
AV-Sync-MOS evaluates the temporal and event-level alignment between audio and visual streams.
PVR-MOS assesses the model’s ability to reflect implicit physical variations described in paired prompts through corresponding changes in audio. Specifically, participants judge how well the audio differences between paired videos correspond to the underlying physical changes implied by the prompts.
In this evaluation, 74 participants rated the extent to which the generated audio captures these subtle yet physically grounded variations.

\subsection{Results}
\label{sec:results}

In this section, we present a comprehensive quantitative evaluation of SOTA T2AV, I2AV, and V2A models on PhyAVBench.
We assess these models across four critical dimensions: generation quality, semantic consistency, audio-visual (AV) synchronization, and physical sensitivity, evaluated via our proposed CPRS metric.

\subsubsection{Basic Evaluation}

\noindent\textbf{1) AV Quality.}
As shown in Tab.~\ref{tab:results}, generation quality metrics (FVD, KVD, and FAD) exhibit different trends across tasks and models.
Open-source T2AV models show competitive visual quality, with JavisDiT++ achieving the lowest FVD, while Sora 2 leads commercial models in KVD. For audio, commercial models dominate, with Wan 2.6 and Kling v2.6 achieving the best FAD.
With image-conditioned I2AV models, both FVD and KVD improve significantly. Ovi achieves the best visual performance, while MOVA attains the best FAD.
V2A models perform best in audio quality. MMAudio achieves the lowest FAD, followed by ThinkSound and HunyuanVideo-Foley.
Overall, the results suggest a trade-off between visual and audio fidelity, and balancing both modalities remains challenging.
\textbf{2) Audio-Visual Semantic Alignment.}
As shown in Tab.~\ref{tab:results}, the evaluated models generally exhibit strong performance on visual semantic consistency metrics.
Among commercial T2AV models, Wan 2.6 achieves the highest TV-IB (0.3434) and CLIP (0.2792), indicating strong alignment of generated visuals with the text prompts. Kling~v2.6 achieves the highest CLAP (0.4252), while Seedance~1.5 Pro achieves the highest TA-IB (0.2258). Within the open-source T2AV group, JavisDiT++ leads in TV-IB, CLIP, CLAP, and TA-IB. MOVA leads the I2AV group in these four metrics, and MMAudio has the highest CLAP and TA-IB among V2A models. The V2A TV-IB and CLIP scores measure the supplied reference videos rather than visual generation by the audio models.
\noindent\textbf{3) Audio-Visual Synchronization.}
As shown in Tab.~\ref{tab:results}, the AV synchronization metrics reveal clear performance differences across models.
Seedance 1.5 Pro leads commercial T2AV models in AV-IB and JavisScore, while Veo~3.1 has the lowest DeSync (0.0160) and Kling~v2.6 attains the highest AVH. Among open-source T2AV models, JavisDiT++ has the highest AV-IB, AVH, and JavisScore, whereas LTX has the lowest DeSync (0.2531).
In I2AV tasks, MOVA leads in synchronization quality, achieving the highest AV-IB, AVH, and Javis scores. However, all models exhibit relatively high DeSync, indicating notable temporal misalignment; LTX performs best in this regard, while MOVA and UniVerse-1 show larger desynchronization.
In V2A tasks, HunyuanVideo-Foley has the highest AV-IB and AVH and the lowest DeSync (0.3072); MMAudio has the second-lowest DeSync (0.3165) with competitive AV metrics.
Overall, conditional settings (I2AV, V2A) provide stronger priors, yet precise AV synchronization, especially reducing temporal desynchronization, remains challenging.

\subsubsection{The APST Challenge}
\noindent\textbf{1) Quantitative Evaluation.}
Despite their high semantic relevance, these models consistently struggle with physical grounding, as reflected by our proposed CPRS.
Even the top-performing commercial T2AV model, Sora~2, achieves a CPRS of only 0.4512, with open-source alternatives like JavisDiT++ scoring lower (0.2844).
Similar bottlenecks are evident in V2A tasks, where the leading model, MMAudio, peaks at a CPRS of 0.4003.
This disparity highlights a critical limitation: while models can synthesize ``semantically plausible'' audio (e.g., pairing a generic impact sound with a hammer strike), they fail to reflect the nuanced, directional acoustic shifts dictated by varying physical properties (e.g., distinguishing the resonance of empty and water-filled plastic bottles).
\noindent\textbf{2) Qualitative Insights.} 
Figure~\ref{fig:qualitative} illustrates the strong correspondence between our CPRS metric and the generated acoustic characteristics. We observe that higher CPRS values, such as those achieved by HunyuanVideo-Foley in the \textit{Helmholtz resonance} scenario, indicate that the model successfully captures the directional trend of key acoustic properties, such as decay time contraction, aligning closely with the physical ground truth.
Conversely, lower CPRS values indicate a distinct lack of physical sensitivity.
In the \textit{sound occlusion} tasks, models like JavisDiT and LTX frequently produce static or contradictory acoustic outputs, failing to exhibit the transitions in pitch or loudness required by the changing physical context. These findings reveal that current models are predominantly semantic-driven, generating sounds based on category rather than physical laws. Our CPRS effectively exposes this gap, distinguishing models that merely synthesize plausible audio types from those capable of genuinely physically grounded generation.

\subsubsection{Error Analysis}

The radar plots in Fig.~\ref{fig:rador} reveal distinct capability patterns across T2AV, I2AV, and V2A models.
In T2AV, Sora2 shows the strongest and most balanced performance, particularly in time/causality and complex coupling, while Wan 2.6 and Seedance1.5Pro are competitive in observer physics and coupling. Open-source models (e.g., JavisDiT, JavisDiT++) lag behind across all dimensions, especially in fluid dynamics and sound propagation.
In I2AV, MOVA excels in fluid-related effects, while LTX and Ovi are relatively balanced; UniVerse-1 performs the worst, especially in temporal reasoning.
In V2A, HunyuanVideo-Foley and MMAudio dominate, with the former strong in temporal and observer-related physics and the latter more balanced overall; ThinkSound and Foleycrafter underperform.
Overall, despite strong performance in certain aspects, all models struggle with fluid dynamics, sound propagation, and complex interactions, indicating limited modeling of underlying audio-physical mechanisms.

\subsubsection{Human Evaluation}

\begin{figure}[!t]
    \centering
    \includegraphics[width=0.95\linewidth]{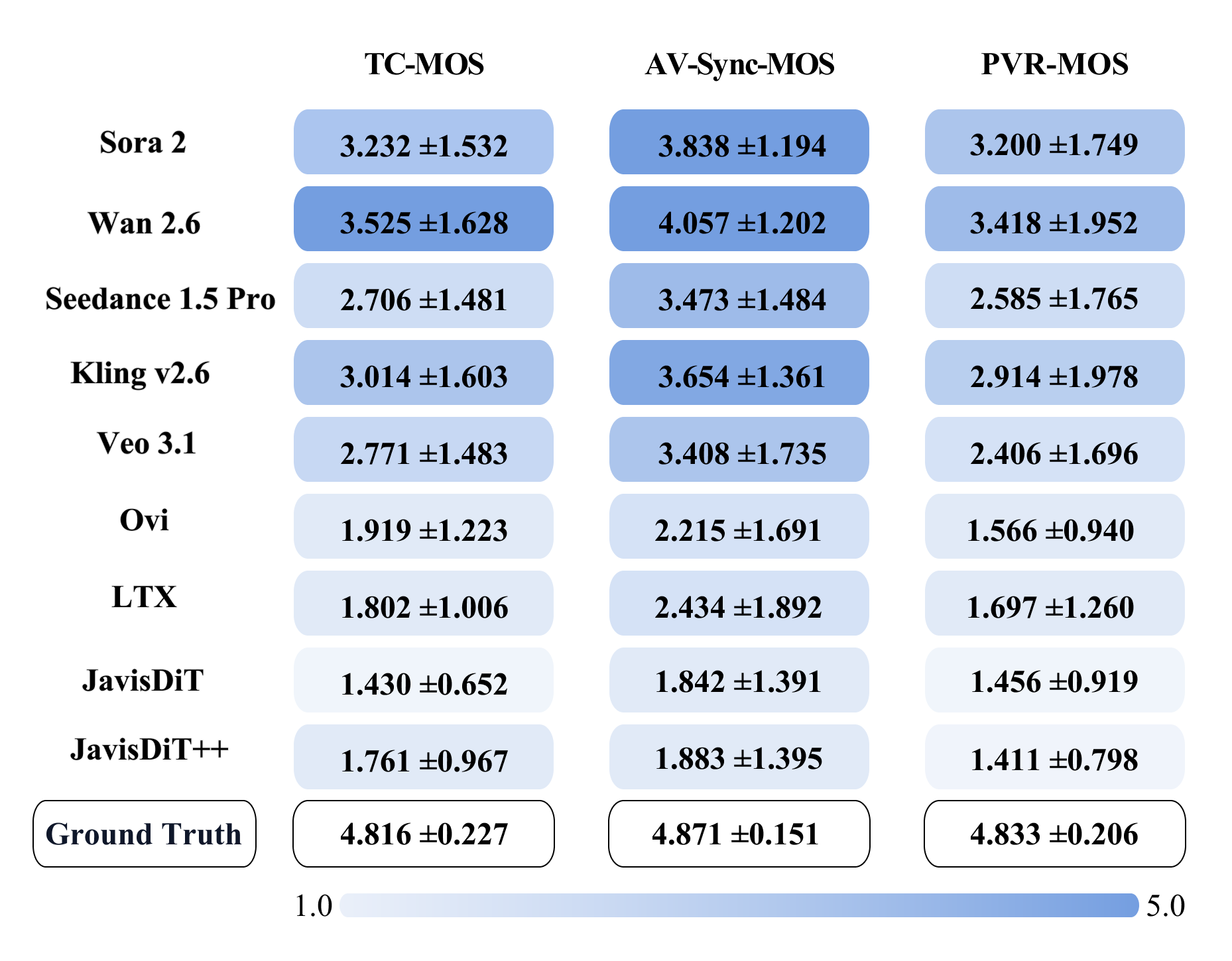}
    \caption{Human evaluation results.}
    \label{fig:mos}
\end{figure}

As shown in Fig.~\ref{fig:mos}, Wan 2.6 emerges as the top-performing model across all three metrics, closely followed by Sora 2 and Kling v2.6, whereas open-source models like JavisDiT and JavisDiT++ struggle significantly with temporal consistency and physical realism. Notably, a substantial ``reality gap'' persists between even the SOTA models and the Ground Truth. This performance deficit is most pronounced in the PVR-MOS metric, underscoring that while recent advancements have improved basic audio-visual synchronization, achieving rigorous physical and acoustic consistency in generated content remains a critical open challenge in the field of audio-visual generation.

\noindent\textbf{Consistency with CPRS.}
As shown in Tab.~\ref{tab:correlation}, CPRS exhibits a higher model-level correlation with human ratings than the other reported metrics (e.g., AVH and Javis scores). For lower-is-better metrics, we negate FAD and DeSync before calculating Pearson correlation so that a positive coefficient consistently indicates agreement with higher PVR-MOS.
This alignment indicates that CPRS effectively captures the subjective notion of physical plausibility and serves as a reliable automated proxy for human perception of audio-physics consistency.

\begin{table}[!t]
\centering
\caption{Pearson correlation (P.C.) across T2AV models, using per-model PVR-MOS means. For lower-is-better metrics, the score is defined as $-\mathrm{DeSync}$ and $-\mathrm{FAD}$.}
\label{tab:correlation}
\renewcommand{\arraystretch}{0.8}
\fontsize{8.5pt}{10pt}\selectfont
\setlength{\tabcolsep}{2pt}
\begin{tabular}{ccccccccc}
\toprule
\textbf{Metric} & $-\mathrm{FAD}$ & TA-IB & CLAP & AV-IB & $-\mathrm{DeSync}$ & AVH & Javis & CPRS \\
\midrule
\textbf{P.C.} & 0.51 & 0.77 & 0.86 & 0.76 & 0.60 & 0.72 & 0.71 & \textbf{0.92} \\
\bottomrule
\end{tabular}

\end{table}

\section{Conclusion}
\label{sec:conclusion}

In this work, we introduce PhyAVBench, a comprehensive benchmark for evaluating the audio-physics grounding capabilities of T2AV, I2AV, and V2A models beyond conventional audio-video synchronization. By incorporating the PhyAV-Sound-11K dataset, paired prompt-based Audio-Physics Sensitivity Tests, and the Contrastive Physical Response Score, PhyAVBench enables systematic and fine-grained assessment of how well generated audio aligns with real-world physical principles. Extensive experiments on 17 state-of-the-art models, together with human studies, show that CPRS exhibits strong alignment with human judgments, while current models still struggle to capture fundamental audio-physical phenomena, revealing a significant gap in physically plausible generation. While PhyAVBench covers a wide range of physical scenarios, it does not yet include certain extreme or hazardous acoustic events. Future work will expand the coverage of physical conditions and explore more advanced evaluation frameworks that incorporate explicit physical laws and formulations, further advancing physically grounded audio-visual generation.

\begin{acks}
This work was supported by the National Natural Science Foundation of China (No. 62471420), GuangDong Basic and Applied Basic Research Foundation (2025A1515012296), and 2025 Tencent AI Lab Rhino-Bird Program.
\end{acks}
    
\bibliographystyle{ACM-Reference-Format}
% \balance
\bibliography{ref}

\newpage
\appendix

\section{Detailed Dataset Information}

\subsection{Audio Physics Coverage}

\begin{figure*}
    \centering
    \includegraphics[width=\linewidth]{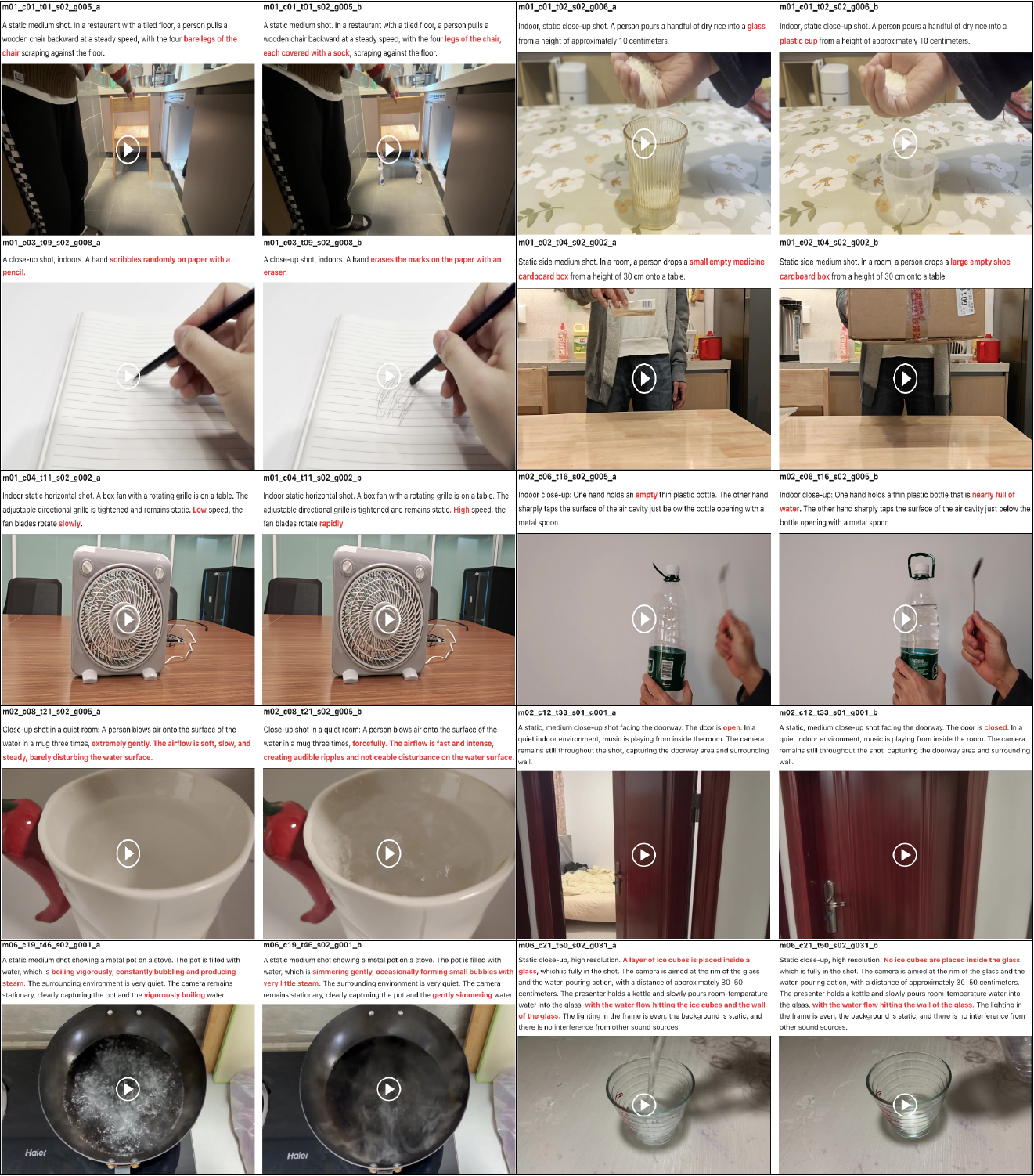}
    \caption{Prompt groups and newly recorded video examples in PhyAV-Sound-11K.}
    \label{fig:prompt_samples}
\end{figure*}

PhyAVBench spans 6 major audio-physics dimensions, each comprising 2--4 subcategories, resulting in a total of 41 fine-grained test points.
These test points range from common acoustic phenomena, such as reverberation and collision sound, to more complex effects, including vortex shedding and Helmholtz resonance, forming a comprehensive evaluation matrix.
Table~\ref{tab:test_points} summarizes all major dimensions, subcategories, and fine-grained test points in PhyAVBench along with representative examples.
Fig.~\ref{fig:prompt_samples} shows some prompts and newly recorded video examples in PhyAV-Sound-11K.

\begin{table*}
% \fontsize{9pt}{\baselineskip}\selectfont
% \setlength{\tabcolsep}{3pt}
\renewcommand{\arraystretch}{0.8}
\centering
\caption{The fine-grained test points and illustrative examples of PhyAVBench.}
\resizebox{1.0\textwidth}{!}{
\begin{threeparttable}
\begin{tabular}{ccccccc}
\toprule
Major Dimension & m Code & Sub Category & c Code & Test Point & t Code & Example \\ 
\midrule
\multirow{13}{*}[-5pt]{\makecell{Sound\\ Source\\ Mechanics}} & \multirow{13}{*}[-5pt]{\makecell{m01}} & \multirow{3}{*}{\makecell{Material\\ Properties}} & \multirow{3}{*}{\makecell{c01}} & Hardness / Damping & t01  & Wrench dropping on concrete vs. carpet \\ 
&&&& Density & t02  & Solid vs. hollow ball impact\\ 
&&&& Surface Texture & t03 & Scrapping wood vs. metal\\ 
\cmidrule(rl){3-7}
&& \multirow{4}{*}{\makecell{Object\\ Geometry}} & \multirow{4}{*}{\makecell{c02}} & Size & t04 & Dropping large vs. small bottoles\\ 
&&&& Shape & t05  & Tapping long tubes vs. flat plates \\ 
&&&& Thickness & t06 & Thin vs. thick steel plates\\ 
&&&& Boundary Conditions & t07 & Rulers: fixed vs. free boundary\\ 
\cmidrule(rl){3-7}
&& \multirow{3}{*}{\makecell{Contact\\ Dynamics}} & \multirow{3}{*}{\makecell{c03}} & Impact Velocity & t08  & Fast vs. slow clapping \\ 
&&  & & Contact Area / Sharpness & t09 & Stiletto vs. flat shoe. \\ 
&&  & & Excitation Continuity & t10 & Tapping vs. dragging \\ 
\cmidrule(rl){3-7}
&& \multirow{3}{*}{\makecell{Mechanical\\ Structures}} & \multirow{3}{*}{\makecell{c04}} & Rotational Speed & t11 & Engine idling vs. accelerating\\ 
&&  & & Looseness & t12 & Tightened vs. loose screws \\ 
&&  & & Tension & t13 & Taut vs. slack rubber bands\\ 
\midrule
\multirow{8}{*}[-5pt]{\makecell{Fluid and\\ Aerodynamics}} & \multirow{8}{*}[-5pt]{\makecell{m02}} & \multirow{2}{*}{\makecell{Volume and Velocity}} & \multirow{2}{*}{\makecell{c05}} & Flow Rate & t14 & Faucet dripping vs. pouring\\ 
&&  & & Fluid Impact / Splash & t15 & Shower head vs. faucet\\ 
\cmidrule(rl){3-7}
&& \multirow{2}{*}{\makecell{Cavity Resonance}} & \multirow{2}{*}{\makecell{c06}} & Helmholtz Resonance & t16 & Blowing empty vs. full bottle\\ 
&&  & & Container Material & t17 & Pouring water into glass vs. plastic cups\\ 
\cmidrule(rl){3-7}
&& \multirow{2}{*}{Viscosity} & \multirow{2}{*}{\makecell{c07}} & Fluid Viscosity & t18 & Pouring water vs. honey\\ 
&&  & & Air Bubble & t19 & Boiling vs. simmering\\ 
\cmidrule(rl){3-7}
&& \multirow{2}{*}{Aerodynamics} & \multirow{2}{*}{\makecell{c08}} & Aeroacoustics / Whoosh & t20 & Fast vs. slow waving a stick\\ 
&&  & & Air Flow Velocity & t21 & Gales vs. breezes. \\ 
\midrule
\multirow{12}{*}[-5pt]{\makecell{Sound\\ Propagation\\ Environment}} & \multirow{12}{*}[-5pt]{\makecell{m03}} & \multirow{4}{*}{Reverberation} & \multirow{4}{*}{\makecell{c09}} & Spatial Volume & t22 & Bathrooms vs. gymnasiums\\ 
&&  & & Surface Absorption & t23 & Carpeted vs. tiled rooms\\ 
&&  & & Echo & t24 & Underground parking vs. room \\ 
&&  & & Space Transition & t25 & Walking out the door\\ 
\cmidrule(rl){3-7}
&& \multirow{2}{*}{\makecell{Wave Behavior}} & \multirow{2}{*}{\makecell{c10}} & Diffraction & t26 & Source behind a corner\\ 
&&  & & Scattering & t27 & Rooms filled with furniture vs. empty rooms. \\ 
\cmidrule(rl){3-7}
&& \multirow{3}{*}{\makecell{Transmission\\ Medium}} & \multirow{3}{*}{\makecell{c11}} & Underwater Acoustics & t28 & Listening from underwater\\ 
&&  & & Solid-Borne Transmission & t29 & Paper cup phone \\ 
&&  & & Dynamic Medium Changes & t31 & Immersing in water\\ 
\cmidrule(rl){3-7}
&& \multirow{2}{*}{\makecell{Occlusion and Insulation}} & \multirow{2}{*}{\makecell{c12}} & Acoustic Leakage & t32 & A closed vs. open door\\ 
&&  & & Sound Insulation & t33 & A closed vs. cracked window\\ 
\midrule
\multirow{3}{*}[-5pt]{\makecell{Observer\\ Physics}} & \multirow{3}{*}[-5pt]{\makecell{m04}} & \multirow{1}{*}{Distance Law} & \multirow{1}{*}{\makecell{c13}} & Inverse Square Law & t34 & Doubling distance halves SPL \\ 
\cmidrule(rl){3-7}
&& \multirow{2}{*}{\makecell{Binaural Effect}} & \multirow{2}{*}{\makecell{c15}} & Horizontal Localization & t38 & Ping pong match\\ 
&&  & & Vertical Localization & t39 & Airplane flying overhead vs. high\\ 
\midrule 
\multirow{2}{*}[-5pt]{\makecell{Time and\\ Causality}} & \multirow{2}{*}[-5pt]{\makecell{m05}} & \multirow{1}{*}{\makecell{Synchronization}} & \multirow{1}{*}{\makecell{c17}} & Vibration Initiation & t42 & Flick a string vs. Covering a vibrating string with a hand\\ 
\cmidrule(rl){3-7}
&& \multirow{1}{*}{Rhythm Consistency} & \multirow{1}{*}{\makecell{c18}} & Periodic \& Aperiodic Motion & t44 & Pendulums vs. Randomly tapping keyboard keys\\ 
\midrule
\multirow{5}{*}[-5pt]{\makecell{Complex\\ Phenomena}} & \multirow{5}{*}[-5pt]{\makecell{m06}} & \multirow{2}{*}{Phase Transition} & \multirow{2}{*}{\makecell{c19}} & Boiling & t46 & Pouring lava into water. \\ 
&&  & & Freezing/Shattering & t47 & Cracking sound of ice cubes in warm liquid. \\ 
\cmidrule(rl){3-7}
&& \multirow{1}{*}{\makecell{Shock Wave}} & \multirow{1}{*}{\makecell{c20}} & Supersonic Speed & t48 & Whip crack (mini sonic boom). \\ 
\cmidrule(rl){3-7}
&& Complex Coupling & c21 & Complex Coupling & t50 & Sound caused by electromagnetic phenomena\\ 
\bottomrule
\end{tabular}
\end{threeparttable}
}
\label{tab:test_points}
\end{table*}

\subsection{Test Point and Duration Distribution}

\begin{figure}[t]
    \centering

    \begin{subfigure}{\linewidth}
        \centering
        \includegraphics[width=\linewidth]{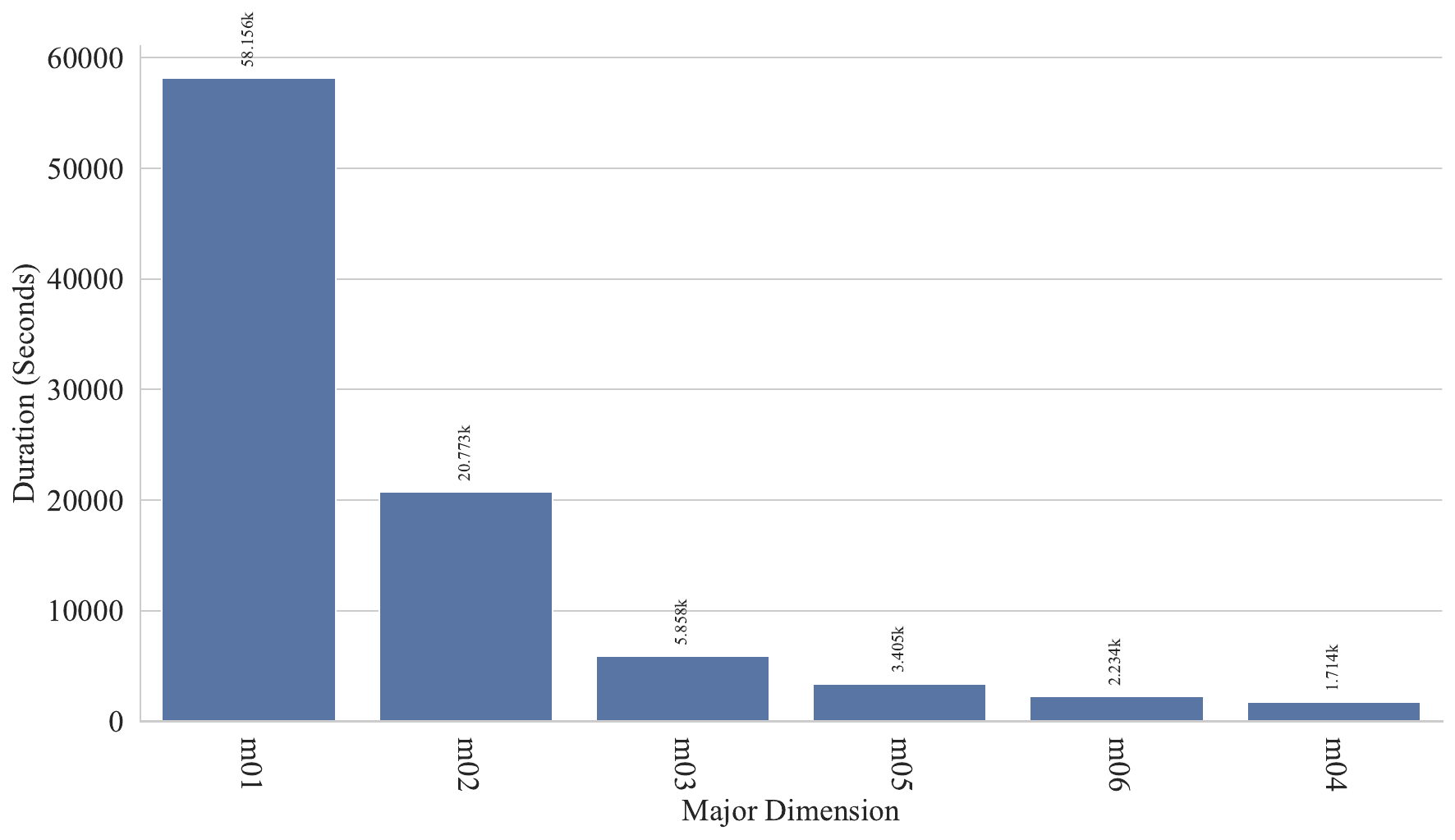}
        \caption{Duration of each major dimension.}
    \end{subfigure}

    \vspace{0.5em}

    \begin{subfigure}{\linewidth}
        \centering
        \includegraphics[width=\linewidth]{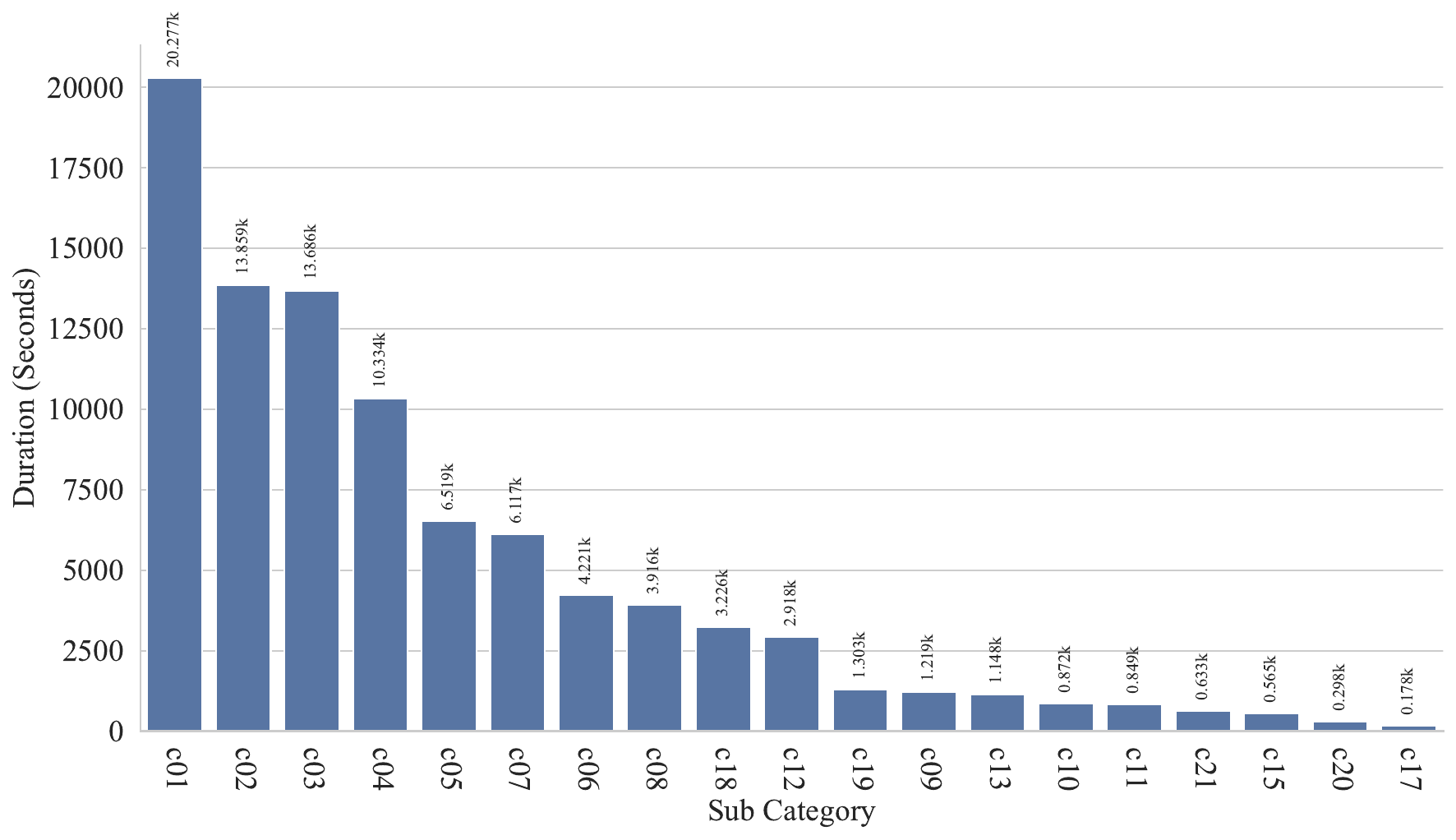}
        \caption{Duration of each subcategory.}
    \end{subfigure}

    \vspace{0.5em}

    \begin{subfigure}{\linewidth}
        \centering
        \includegraphics[width=\linewidth]{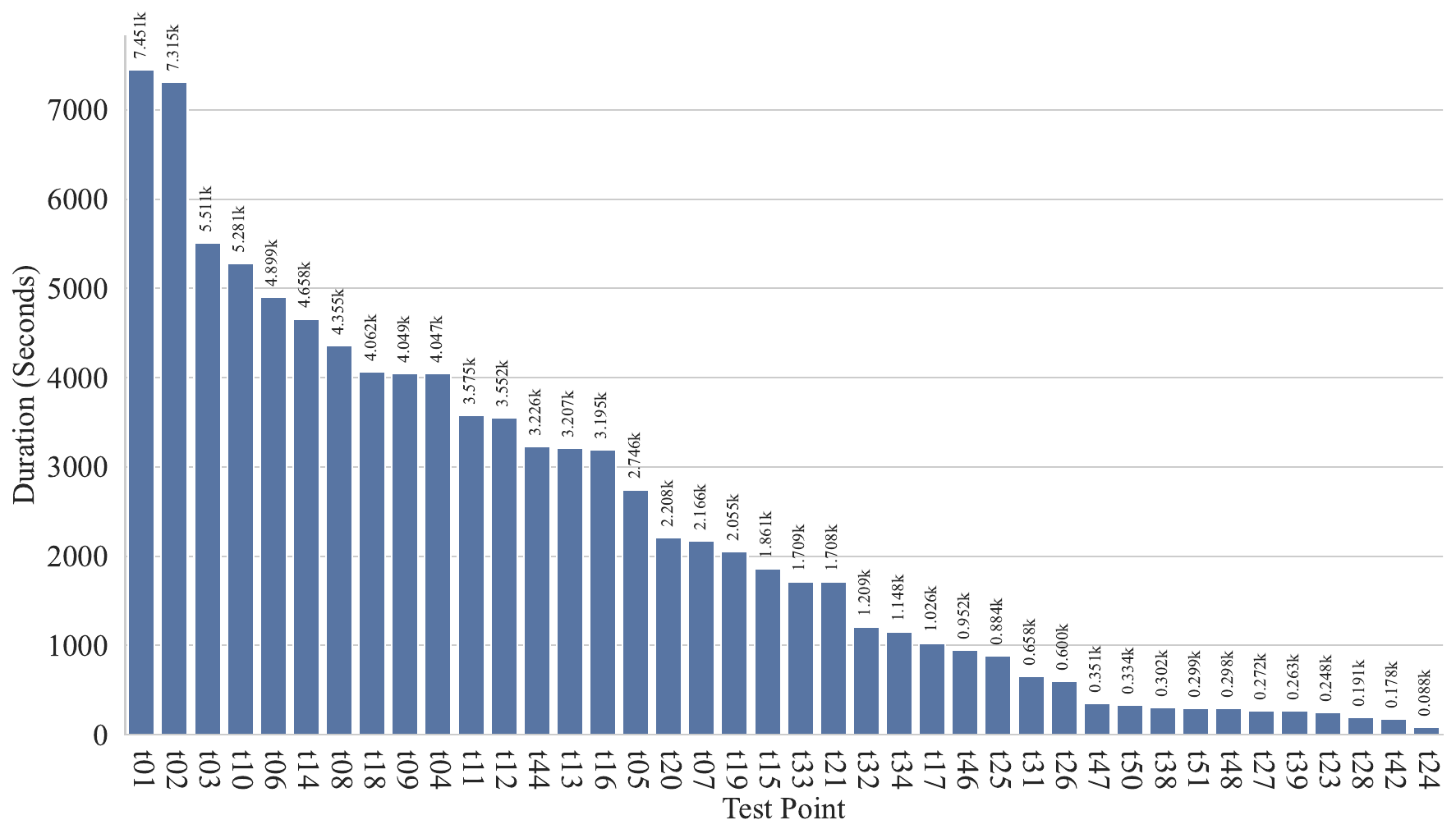}
        \caption{Duration of each test point.}
    \end{subfigure}

    \caption{The duration distribution of PhyAV-Sound-11K.}
    \label{fig:duration_stats}
\end{figure}

As shown in Fig.~\ref{fig:duration_stats}, the phyav-sound-11k dataset presents a characteristic long-tail duration distribution across its three hierarchical levels: \textit{Major Dimension}, \textit{Sub Category}, and \textit{Test Point}.
At the macro level, a substantial portion of the data is concentrated in the major dimension \textit{m01-Sound Source Mechanics}, which contains 63.284k seconds of audio, followed by \textit{m02-Fluid and Aerodynamics} with 16.009k seconds, while \textit{m04-Observer Physics} has the shortest duration at only 1.148k seconds.
This distribution naturally extends to the subcategory level, where \textit{c01-Material Properties} accounts for 21.952k seconds, with other major subcategories such as \textit{c03-Contact Dynamics} (13.686k seconds), \textit{c02-Object Geometry} (12.477k seconds), and \textit{c04-Mechanical Structures} (10.334k seconds) also contributing significant amounts of data.
Additional subcategories cover a diverse range of less frequent scenarios, collectively forming the long-tail portion of the dataset.

At the most fine-grained test point level, representative scenarios such as t01 (7.451k seconds) and t02 (7.315k seconds) account for a considerable share of the total duration, while the remaining data spans a wide spectrum of less common test points, including cases like t24 with 0.088k seconds.

This distribution reflects the inherent characteristics of real-world audio phenomena. In everyday environments, a large proportion of observable events naturally cluster around a limited set of common scenarios, leading to higher data availability for corresponding test points. In contrast, rarer or more specialized acoustic events tend to occur less frequently and are often more challenging to capture or reproduce during data collection, resulting in comparatively smaller data volumes for those test points.

\subsection{Difficulty Distribution}

\begin{figure}[ht]
    \centering

    \begin{subfigure}{\linewidth}
        \centering
        \includegraphics[width=\linewidth]{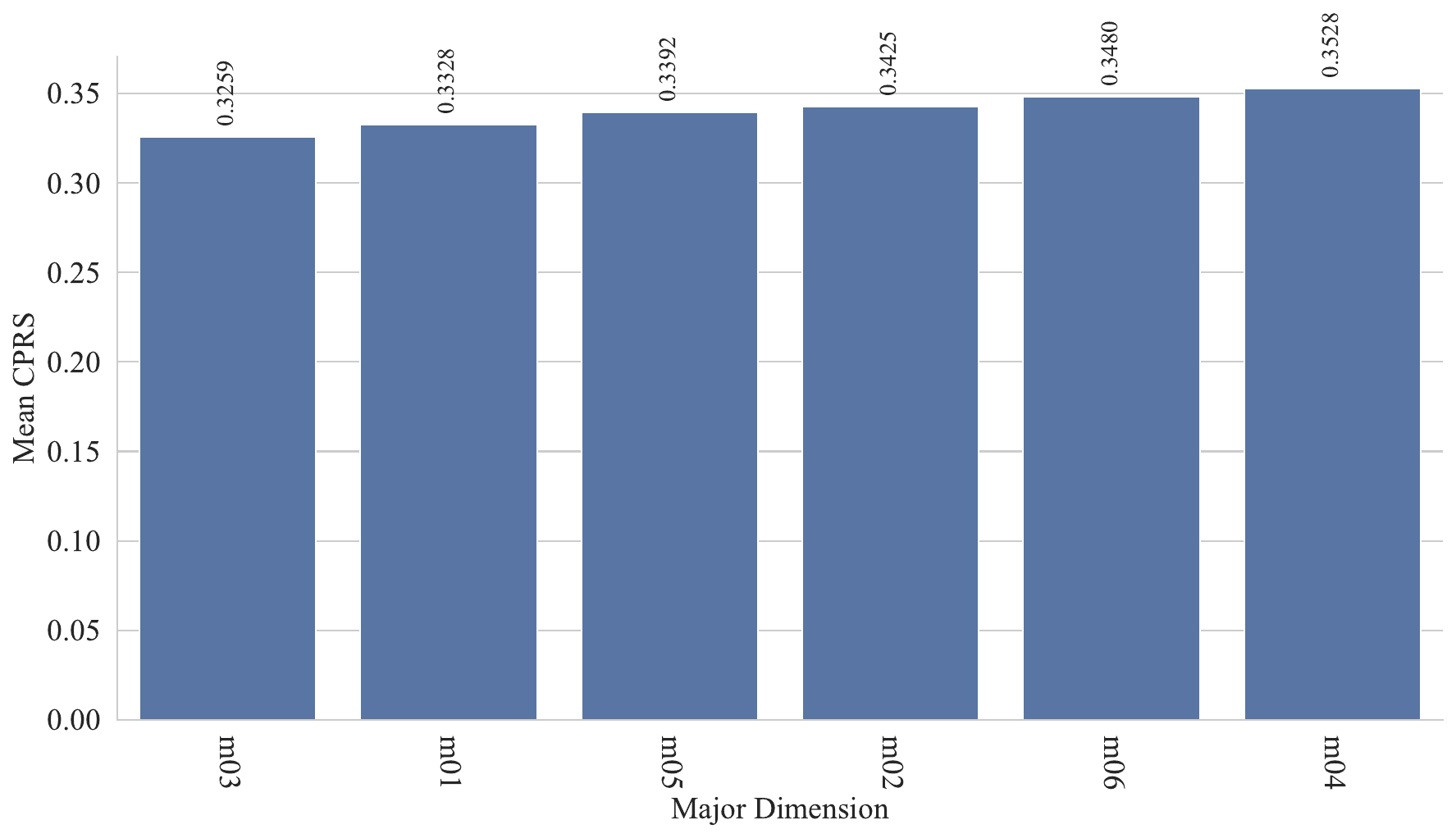}
        \caption{Average CPRS of each major dimension.}
    \end{subfigure}

    \vspace{0.5em}

    \begin{subfigure}{\linewidth}
        \centering
        \includegraphics[width=\linewidth]{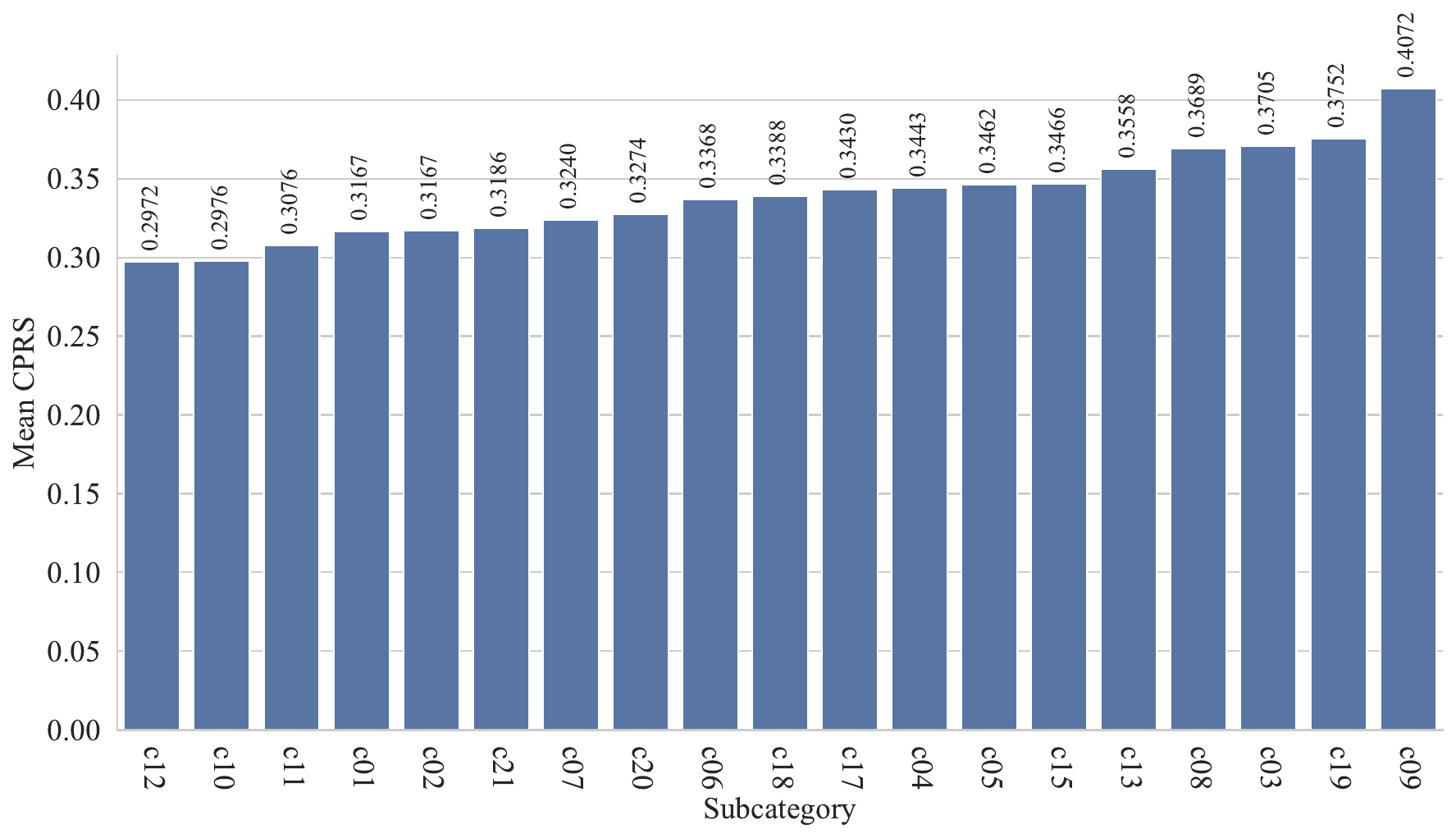}
        \caption{Average CPRS of each subcategory.}
    \end{subfigure}

    \vspace{0.5em}

    \begin{subfigure}{\linewidth}
        \centering
        \includegraphics[width=\linewidth]{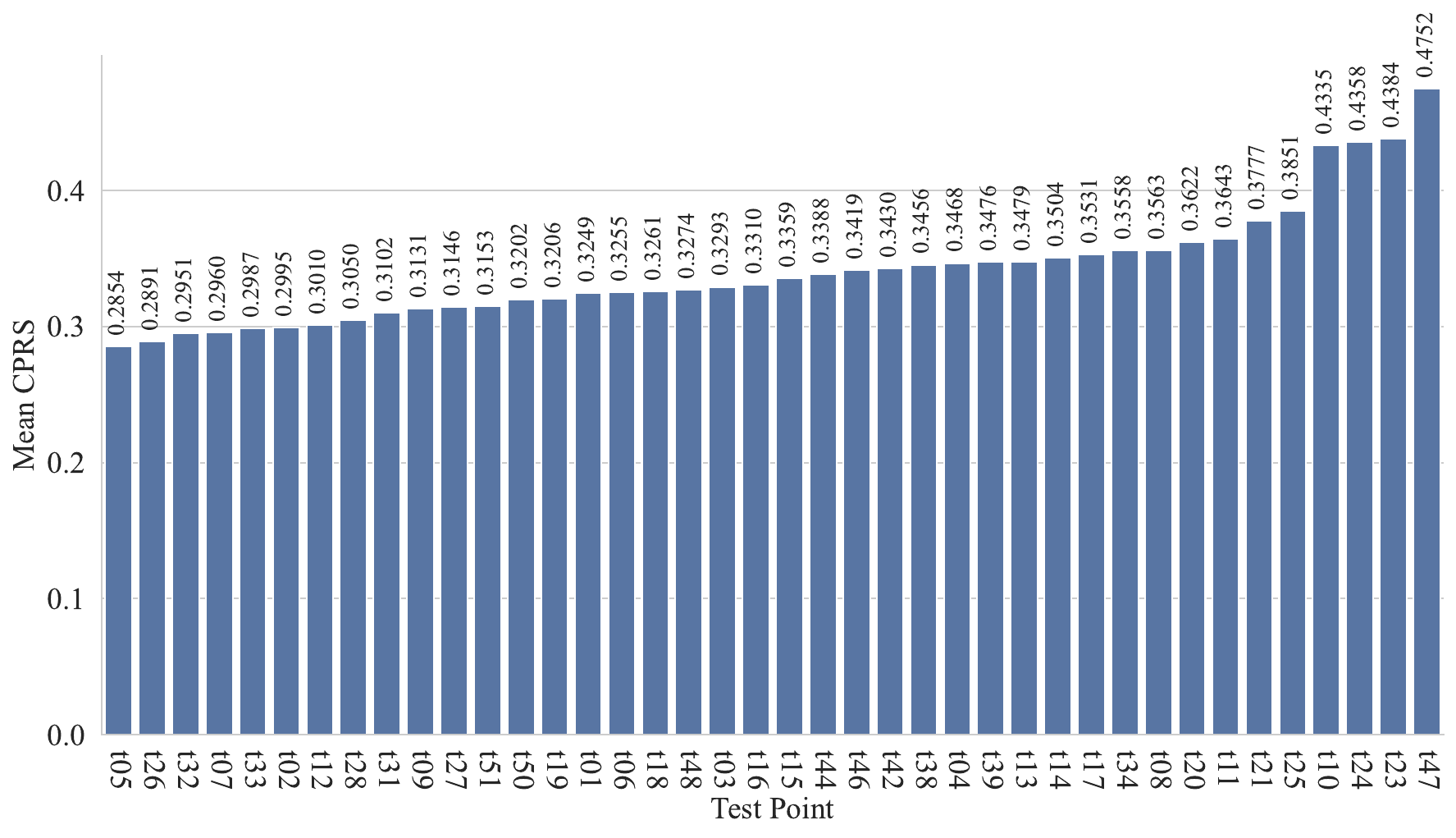}
        \caption{Average CPRS of each test point.}
    \end{subfigure}

    \caption{The difficulty distribution of PhyAV-Sound-11K.}
    \label{fig:difficulty_stats}
\end{figure}

In this subsection, we report the difficulty distribution of PhyAV-Sound-11K. We utilize all 4,404 original LAION-CLAP CPRS scores, each calculated for one A/B prompt pair from the 8,808 available generated video samples across the models \cite{sora2,veo3.1,seedance1.5pro,Kling-V2-6,Wan2.6,liu2026javisdit++,liu2025javisdit,low2025ovi,hacohen2026ltx,wang2025universe,team2026mova,low2025ovi,hacohen2026ltx,zhang2024foleycrafter,shan2025hunyuanvideo,cheng2025mmaudio,liu2025thinksound} evaluated in Table~1 of the main paper.
As shown in Fig.~\ref{fig:difficulty_stats}, the difficulty distribution of the PhyAVBench dataset is structured as a three-tier hierarchy with increasing granularity. Performance is evaluated using the Mean CPRS, where lower values indicate greater task complexity.

At the \textit{Major Dimension} level, performance is relatively concentrated within a narrow range from 0.3259 to 0.3528. Among these, \textit{m03-Sound Propagation Environment} corresponds to comparatively higher complexity with a score of 0.3259, while \textit{m04-Observer Physics} achieves a higher score of 0.3528, indicating relatively stronger model performance.

Moving to the \textit{Subcategory} level, a broader range of performance emerges across the 19 subcategories represented by the available model outputs. Subcategories such as \textit{c12-Occlusion and Insulation} (0.2972) and c10 (0.2976) are associated with more challenging scenarios, whereas \textit{c09-Reverberation} (0.4072) and \textit{c19-Phase Transition} (0.3752) correspond to cases where models achieve higher scores.

At the most fine-grained Test Point level, the benchmark captures a wide spectrum of task characteristics. Test points such as \textit{t05-Shape} (0.2854) and \textit{t26-Diffraction} (0.2891) reflect relatively complex cases, while others like \textit{t47-Freezing/Shattering} reach higher performance levels, with a Mean CPRS of 0.4752.

Overall, this hierarchical organization highlights that while high-level dimensions exhibit relatively stable performance, finer-grained test points provide a more detailed view of variability, enabling comprehensive assessment across diverse audio-physical scenarios.

\section{Additional Experiments}

\subsection{Ablation Study on Embedding Models}

To evaluate the robustness of the proposed CPRS, we compute the rankings using three different embedding models: LAION-CLAP~\cite{wu2023large}, ImageBind~\cite{girdhar2023imagebind}, and CAV-MAE-Sync~\cite{araujo2025cav}.
As shown across the three tables, the overall ranking patterns remain largely consistent under different embedding backbones, indicating that CPRS is not overly sensitive to a specific representation space.
In particular, strong-performing models such as MOVA (for I2AV), Sora 2 (for T2AV), and MMAudio (for V2A) consistently rank near the top across all embedding models, despite minor fluctuations in absolute CPRS values.
Similarly, lower-ranked models tend to maintain their relative positions, suggesting stable comparative behavior.

While there are small variations in exact CPRS scores and standard deviations across embedding models, these differences do not significantly alter the overall ordering within each task category.
For example, models like LTX and Ovi exhibit moderate rank shifts but remain within a comparable performance tier across all three embeddings.

Table~\ref{tab:corr_embed_comparison} reports the Pearson correlation between CPRS (computed with different embedding models) and human evaluation scores (PVR-MOS). Overall, all three embedding backbones exhibit a strong positive correlation with human judgments, with coefficients above 0.89, indicating that CPRS is highly aligned with human perception regardless of the underlying embedding space.
Among them, ImageBind achieves the highest correlation (0.9457), suggesting that it provides the most human-consistent representation for measuring audio-visual quality in this setting. LAION-CLAP also shows a comparably strong correlation (0.9230), while CAV-MAE-Sync, although slightly lower (0.8927), still maintains a high level of agreement.

These results further reinforce the robustness of CPRS: even when instantiated with different embedding models trained on diverse objectives, it consistently correlates well with human evaluation. The relatively small performance gap across embeddings suggests that CPRS captures intrinsic properties of audio-visual grounding rather than relying on a specific feature extractor.

\begin{table*}[htbp]
\label{tab:cprs_clap}
\renewcommand{\arraystretch}{0.89}
  \centering
  \caption{CPRS ranking based on the LAION-CLAP embedding model.}
  \small
  \begin{tabular}{lllcccc}
    \toprule
    Rank & Task & Model & CPRS $\pm$ std & Cos $\pm$ std & $p$ $\pm$ std & $f(p)$ $\pm$ std \\
    \midrule
    01 & I2AV & MOVA & \textbf{0.3613 $\pm$ 0.1891} & \textbf{0.0853 $\pm$ 0.2335} & \textbf{0.172 $\pm$ 0.526} & \textbf{0.1799 $\pm$ 0.2841} \\
    02 & I2AV & LTX & 0.3285 $\pm$ 0.1544 & 0.0652 $\pm$ 0.1712 & 0.148 $\pm$ 0.579 & 0.1244 $\pm$ 0.2452 \\
    03 & I2AV & Ovi & 0.3198 $\pm$ 0.1429 & 0.0580 $\pm$ 0.1840 & 0.104 $\pm$ 0.386 & 0.1106 $\pm$ 0.2177 \\
    04 & I2AV & UniVerse-1 & 0.2729 $\pm$ 0.0812 & 0.0084 $\pm$ 0.1172 & 0.018 $\pm$ 0.283 & 0.0416 $\pm$ 0.1236 \\
    \midrule
    01 & T2AV & Sora 2 & \textbf{0.4512 $\pm$ 0.2149} & \textbf{0.1694 $\pm$ 0.2213} & \textbf{0.389 $\pm$ 0.580} & \textbf{0.3177 $\pm$ 0.3432} \\
    02 & T2AV & Wan 2.6 & 0.4120 $\pm$ 0.2173 & 0.1209 $\pm$ 0.2261 & 0.275 $\pm$ 0.643 & 0.2635 $\pm$ 0.3455 \\
    03 & T2AV & Seedance-1-5-Pro & 0.4044 $\pm$ 0.2009 & 0.1267 $\pm$ 0.2031 & 0.280 $\pm$ 0.492 & 0.2455 $\pm$ 0.3213 \\
    04 & T2AV & Kling-v2-6 & 0.3718 $\pm$ 0.1803 & 0.1110 $\pm$ 0.2011 & 0.252 $\pm$ 0.464 & 0.1881 $\pm$ 0.2815 \\
    05 & T2AV & Veo3.1 & 0.3563 $\pm$ 0.1791 & 0.0898 $\pm$ 0.2037 & 0.149 $\pm$ 0.442 & 0.1676 $\pm$ 0.2733 \\
    06 & T2AV & Ovi & 0.3290 $\pm$ 0.1608 & 0.0629 $\pm$ 0.2039 & 0.128 $\pm$ 0.462 & 0.1264 $\pm$ 0.2456 \\
    07 & T2AV & LTX & 0.3235 $\pm$ 0.1560 & 0.0549 $\pm$ 0.1814 & 0.135 $\pm$ 0.495 & 0.1196 $\pm$ 0.2443 \\
    08 & T2AV & JavisDit & 0.2938 $\pm$ 0.1117 & 0.0472 $\pm$ 0.1634 & 0.080 $\pm$ 0.334 & 0.0639 $\pm$ 0.1689 \\
    09 & T2AV & JavisDit++ & 0.2844 $\pm$ 0.1026 & 0.0289 $\pm$ 0.1692 & 0.037 $\pm$ 0.343 & 0.0545 $\pm$ 0.1471 \\
    \midrule
    01 & V2A & MMAudio & \textbf{0.4003 $\pm$ 0.2024} & \textbf{0.1533 $\pm$ 0.2298} & \textbf{0.286 $\pm$ 0.478} & \textbf{0.2240 $\pm$ 0.3106} \\
    02 & V2A & Hunyuanvideo-Foley & 0.3896 $\pm$ 0.2006 & 0.1238 $\pm$ 0.2130 & 0.247 $\pm$ 0.494 & 0.2173 $\pm$ 0.3178 \\
    03 & V2A & Thinksound & 0.3186 $\pm$ 0.1366 & 0.0616 $\pm$ 0.1704 & 0.115 $\pm$ 0.377 & 0.1063 $\pm$ 0.2097 \\
    04 & V2A & Foleycrafter & 0.3019 $\pm$ 0.1194 & 0.0500 $\pm$ 0.1638 & 0.085 $\pm$ 0.302 & 0.0788 $\pm$ 0.1839 \\
    \bottomrule
  \end{tabular}
\end{table*}

\begin{table*}[htbp]
\label{tab:cprs_imagebind}
\renewcommand{\arraystretch}{0.89}
  \centering
  \caption{CPRS ranking based on the ImageBind embedding model.}
  \small
  \begin{tabular}{lllcccc}
    \toprule
    Rank & Task & Model & CPRS $\pm$ std & Cos $\pm$ std & $p$ $\pm$ std & $f(p)$ $\pm$ std \\
    \midrule
    01 & I2AV & MOVA & \textbf{0.3688 $\pm$ 0.2001} & \textbf{0.0501 $\pm$ 0.1925} & \textbf{0.128 $\pm$ 0.712} & \textbf{0.2125 $\pm$ 0.3274} \\
    02 & I2AV & LTX & 0.3125 $\pm$ 0.1422 & 0.0203 $\pm$ 0.1297 & 0.062 $\pm$ 0.485 & 0.1149 $\pm$ 0.2377 \\
    03 & I2AV & Ovi & 0.3005 $\pm$ 0.1269 & 0.0238 $\pm$ 0.1322 & 0.050 $\pm$ 0.394 & 0.0892 $\pm$ 0.2070 \\
    04 & I2AV & UniVerse-1 & 0.2844 $\pm$ 0.1038 & 0.0061 $\pm$ 0.1066 & 0.010 $\pm$ 0.378 & 0.0657 $\pm$ 0.1711 \\
    \midrule
    01 & T2AV & Sora 2 & \textbf{0.4288 $\pm$ 0.2214} & \textbf{0.1148 $\pm$ 0.1883} & \textbf{0.350 $\pm$ 0.756} & \textbf{0.3003 $\pm$ 0.3716} \\
    02 & T2AV & Wan 2.6 & 0.3850 $\pm$ 0.1961 & 0.0766 $\pm$ 0.2145 & 0.286 $\pm$ 0.947 & 0.2317 $\pm$ 0.3206 \\
    03 & T2AV & Kling-v2-6 & 0.3806 $\pm$ 0.2011 & 0.0665 $\pm$ 0.1951 & 0.237 $\pm$ 0.716 & 0.2280 $\pm$ 0.3288 \\
    04 & T2AV & Seedance-1-5-Pro & 0.3756 $\pm$ 0.1869 & 0.0700 $\pm$ 0.1874 & 0.246 $\pm$ 0.566 & 0.2162 $\pm$ 0.3039 \\
    05 & T2AV & Veo3.1 & 0.3365 $\pm$ 0.1631 & 0.0566 $\pm$ 0.1679 & 0.148 $\pm$ 0.559 & 0.1446 $\pm$ 0.2686 \\
    06 & T2AV & LTX & 0.3110 $\pm$ 0.1383 & 0.0306 $\pm$ 0.1358 & 0.079 $\pm$ 0.480 & 0.1067 $\pm$ 0.2289 \\
    07 & T2AV & Ovi & 0.3066 $\pm$ 0.1304 & 0.0310 $\pm$ 0.1417 & 0.083 $\pm$ 0.444 & 0.0976 $\pm$ 0.2103 \\
    08 & T2AV & JavisDit++ & 0.2833 $\pm$ 0.1067 & 0.0147 $\pm$ 0.1480 & 0.015 $\pm$ 0.435 & 0.0593 $\pm$ 0.1634 \\
    09 & T2AV & JavisDit & 0.2814 $\pm$ 0.0895 & 0.0255 $\pm$ 0.1360 & 0.037 $\pm$ 0.352 & 0.0500 $\pm$ 0.1379 \\
    \midrule
    01 & V2A & MMAudio & \textbf{0.3741 $\pm$ 0.1894} & \textbf{0.0882 $\pm$ 0.1882} & \textbf{0.217 $\pm$ 0.516} & \textbf{0.2040 $\pm$ 0.3043} \\
    02 & V2A & Hunyuanvideo-Foley & 0.3636 $\pm$ 0.1855 & 0.0654 $\pm$ 0.1924 & 0.192 $\pm$ 0.587 & 0.1945 $\pm$ 0.2967 \\
    03 & V2A & Thinksound & 0.3163 $\pm$ 0.1491 & 0.0315 $\pm$ 0.1657 & 0.082 $\pm$ 0.448 & 0.1168 $\pm$ 0.2367 \\
    04 & V2A & Foleycrafter & 0.3093 $\pm$ 0.1345 & 0.0431 $\pm$ 0.1649 & 0.091 $\pm$ 0.382 & 0.0970 $\pm$ 0.2124 \\
    \bottomrule
  \end{tabular}
\end{table*}

\begin{table*}[htbp]
\label{tab:cprs_cav_mae_sync}
\renewcommand{\arraystretch}{0.89}
  \centering
  \caption{CPRS ranking based on the CAV-MAE-Sync embedding model.}
  \small
  \begin{tabular}{lllcccc}
    \toprule
    Rank & Task & Model & CPRS $\pm$ std & Cos $\pm$ std & $p$ $\pm$ std & $f(p)$ $\pm$ std \\
    \midrule
    01 & I2AV & MOVA & \textbf{0.3677 $\pm$ 0.1972} & \textbf{0.0879 $\pm$ 0.2245} & \textbf{0.197 $\pm$ 0.666} & \textbf{0.1914 $\pm$ 0.3075} \\
    02 & I2AV & LTX & 0.3098 $\pm$ 0.1307 & 0.0602 $\pm$ 0.1834 & 0.096 $\pm$ 0.395 & 0.0895 $\pm$ 0.1961 \\
    03 & I2AV & Ovi & 0.2983 $\pm$ 0.0935 & 0.0760 $\pm$ 0.1701 & 0.085 $\pm$ 0.251 & 0.0586 $\pm$ 0.1318 \\
    04 & I2AV & UniVerse-1 & 0.2720 $\pm$ 0.0794 & 0.0226 $\pm$ 0.1375 & 0.014 $\pm$ 0.218 & 0.0326 $\pm$ 0.1173 \\
    \midrule
    01 & T2AV & Wan 2.6 & \textbf{0.3595 $\pm$ 0.1690} & \textbf{0.1179 $\pm$ 0.2207} & \textbf{0.162 $\pm$ 0.409} & \textbf{0.1600 $\pm$ 0.2519} \\
    02 & T2AV & Kling-v2-6 & 0.3482 $\pm$ 0.1356 & 0.1347 $\pm$ 0.2054 & 0.185 $\pm$ 0.315 & 0.1291 $\pm$ 0.1962 \\
    03 & T2AV & Sora 2 & 0.3479 $\pm$ 0.1498 & 0.1347 $\pm$ 0.1913 & 0.170 $\pm$ 0.356 & 0.1285 $\pm$ 0.2337 \\
    04 & T2AV & Seedance-1-5-Pro & 0.3302 $\pm$ 0.1328 & 0.1147 $\pm$ 0.1893 & 0.145 $\pm$ 0.303 & 0.1030 $\pm$ 0.2138 \\
    05 & T2AV & LTX & 0.3052 $\pm$ 0.1243 & 0.0573 $\pm$ 0.1835 & 0.066 $\pm$ 0.389 & 0.0817 $\pm$ 0.1803 \\
    06 & T2AV & Ovi & 0.2991 $\pm$ 0.1010 & 0.0764 $\pm$ 0.1853 & 0.083 $\pm$ 0.258 & 0.0600 $\pm$ 0.1388 \\
    07 & T2AV & Veo3.1 & 0.2841 $\pm$ 0.0588 & 0.0760 $\pm$ 0.1630 & 0.058 $\pm$ 0.201 & 0.0302 $\pm$ 0.0521 \\
    08 & T2AV & JavisDit & 0.2831 $\pm$ 0.0820 & 0.0597 $\pm$ 0.1630 & 0.050 $\pm$ 0.204 & 0.0363 $\pm$ 0.1162 \\
    09 & T2AV & JavisDit++ & 0.2716 $\pm$ 0.0737 & 0.0265 $\pm$ 0.1641 & -0.003 $\pm$ 0.295 & 0.0300 $\pm$ 0.0946 \\
    \midrule
    01 & V2A & MMAudio & \textbf{0.4044 $\pm$ 0.2000} & \textbf{0.1626 $\pm$ 0.2128} & \textbf{0.313 $\pm$ 0.507} & \textbf{0.2274 $\pm$ 0.3193} \\
    02 & V2A & Hunyuanvideo-Foley & 0.3981 $\pm$ 0.1997 & 0.1398 $\pm$ 0.2279 & 0.298 $\pm$ 0.582 & 0.2263 $\pm$ 0.3134 \\
    03 & V2A & Foleycrafter & 0.3163 $\pm$ 0.1422 & 0.0638 $\pm$ 0.1806 & 0.099 $\pm$ 0.374 & 0.1007 $\pm$ 0.2215 \\
    04 & V2A & Thinksound & 0.3150 $\pm$ 0.1310 & 0.0649 $\pm$ 0.1906 & 0.105 $\pm$ 0.455 & 0.0975 $\pm$ 0.1994 \\
    \bottomrule
  \end{tabular}
\end{table*}

\begin{table}[!t]
\centering
\caption{Pearson correlation (P.C.) between CPRS metrics based on three embedding models and the PVR-MOS.}
\label{tab:corr_embed_comparison}
\begin{tabular}{cccc}
\toprule
\textbf{Metric} & LAION-CLAP & ImageBind & CAV-MAE-Sync \\
\midrule
\textbf{P.C.} & 0.9230 & \textbf{0.9457} & 0.8927 \\
\bottomrule
\end{tabular}

\end{table}

\subsection{Details of Human Evaluation}

\begin{figure}
    \centering
    \includegraphics[width=\linewidth]{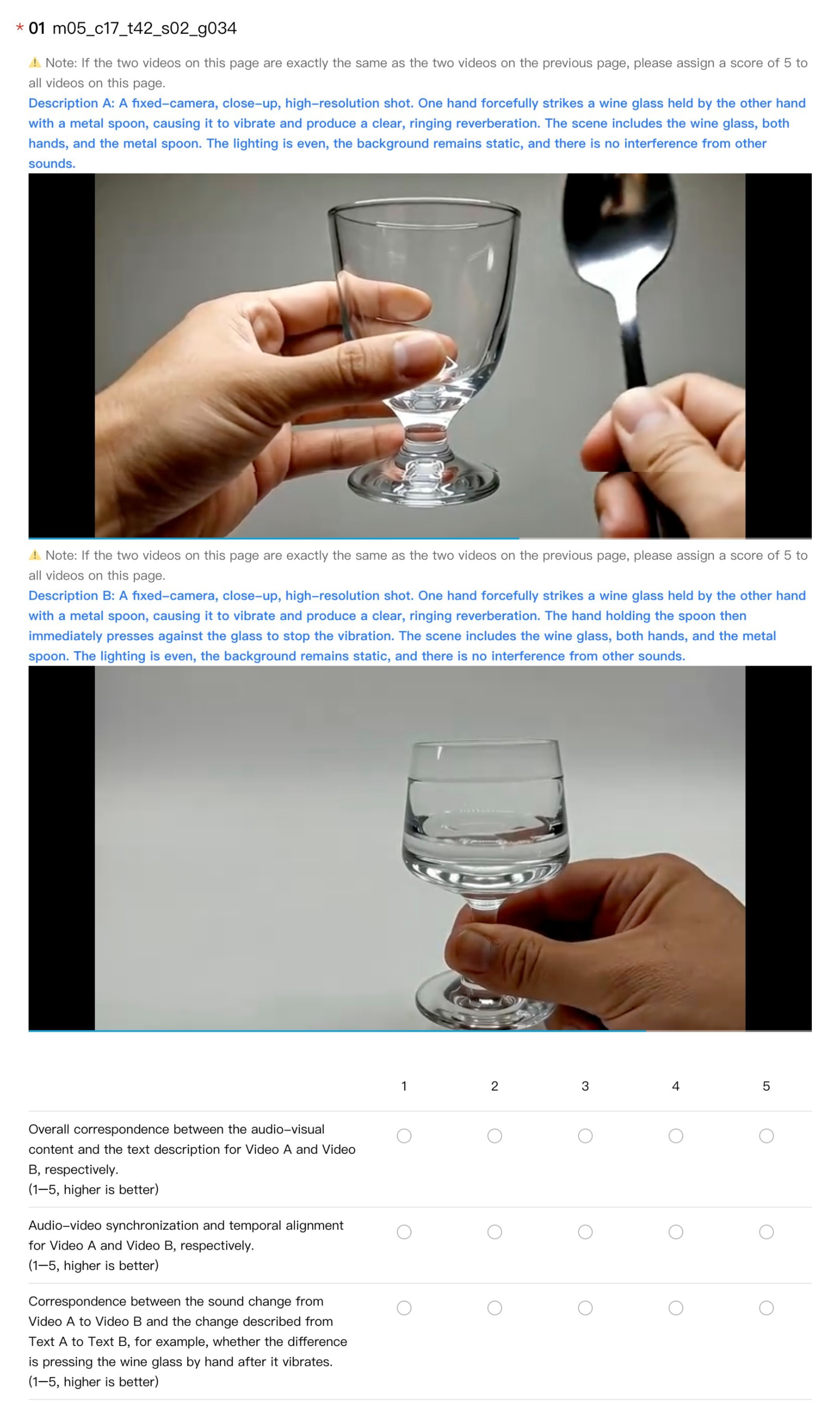}
    \caption{The example questionnaire for human evaluation.}
    \label{fig:questionnaire}
\end{figure}

We conduct a comprehensive human evaluation to validate the effectiveness of CPRS. Specifically, we sample 20 prompt groups, each evaluated across 10 T2AV models (including the ground truth)~\cite{sora2,veo3.1,seedance1.5pro,Kling-V2-6,Wan2.6,liu2026javisdit++,liu2025javisdit,low2025ovi,hacohen2026ltx}, ensuring diverse and representative coverage of test points.
As shown in Fig.~\ref{fig:questionnaire}, for each model output, annotators are asked to answer three questions addressing \textit{semantic consistency}, \textit{audio-visual synchronization}, and \textit{responsiveness to physical changes}.
In total, this results in 600 evaluation questions. The study involves 74 human participants, and after quality control, we retain 419 valid responses. The results show that CPRS exhibits strong agreement with human judgments, demonstrating its reliability as an automatic metric for evaluating audio-visual generation quality.

\begin{figure*}[t]
    \centering
    \includegraphics[width=0.8\linewidth]{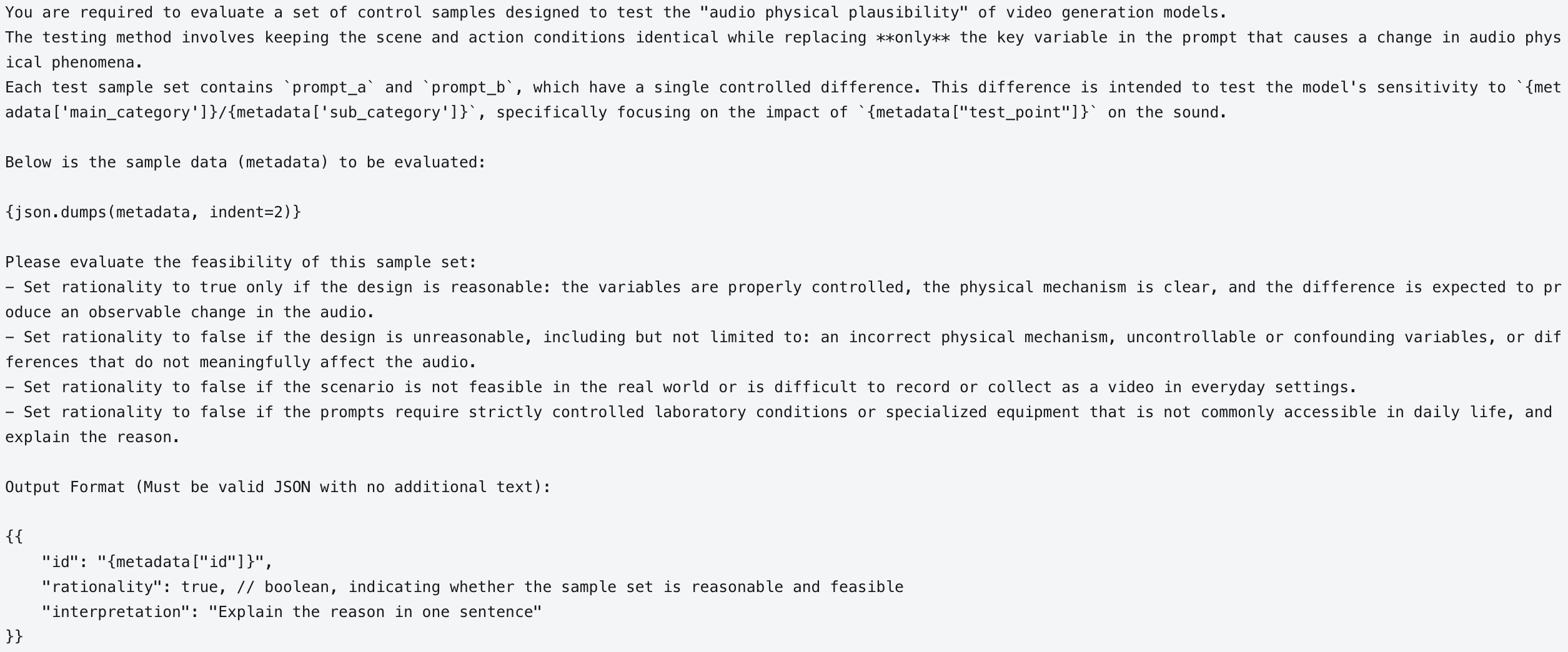}
    \caption{The prompt for cross-model validation.}
    \label{fig:prompt_validation}
\end{figure*}

\begin{figure*}[t]
    \centering
    \includegraphics[width=\linewidth]{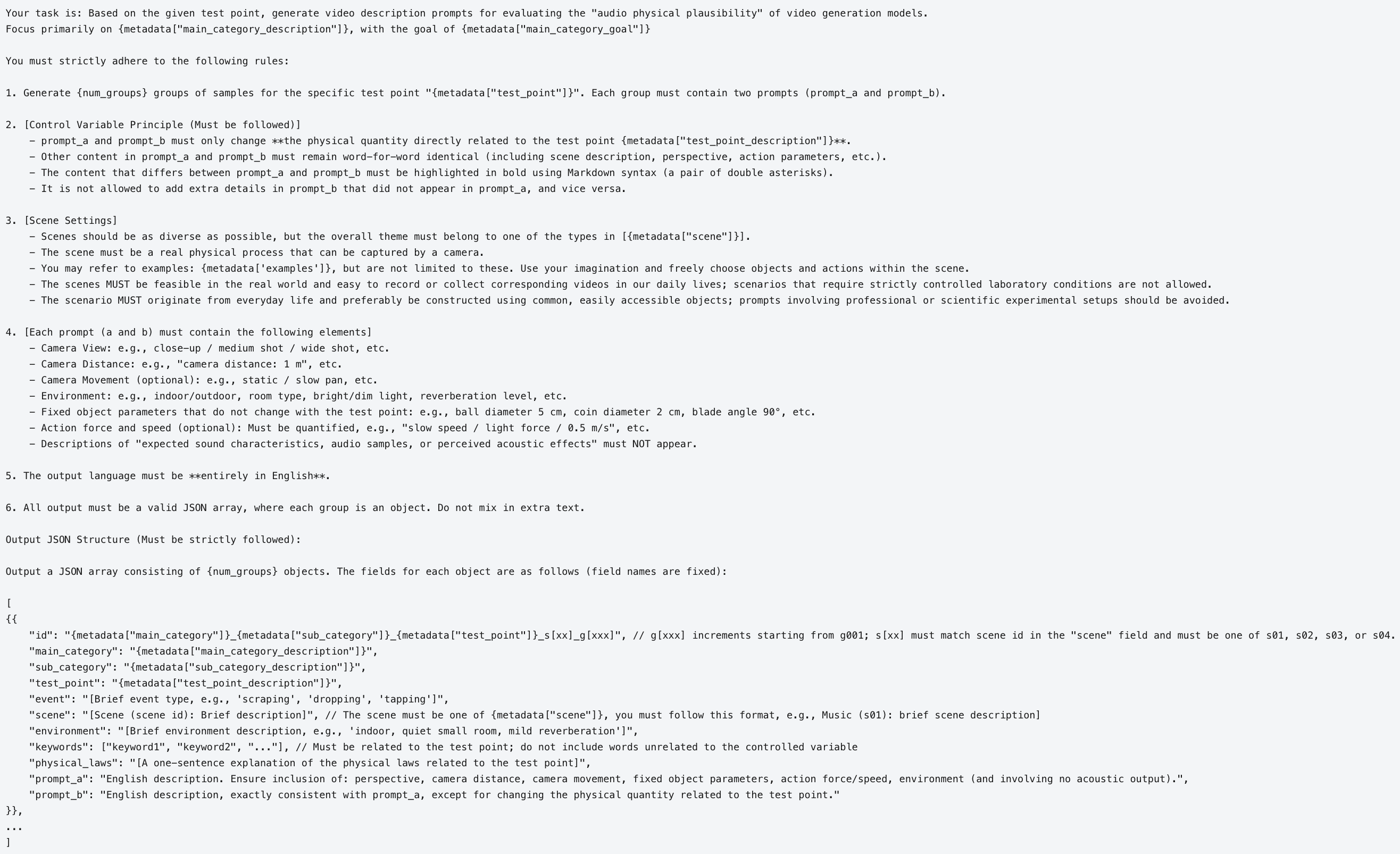}
    \caption{The prompt used to generate prompt groups based on the taxonomy of PhyAVBench.}
    \label{fig:prompt_generation}
\end{figure*}

\section{Data Curation Prompts}

\subsection{Audio-Physics Knowledge Survey}

To ensure comprehensive coverage of real-world audio-physics phenomena, we first conduct an audio-physics knowledge survey through a human--LLM collaborative process. Specifically, domain experts and GPT-5.1 engage in iterative brainstorming and discussion, where the model proposes diverse physical scenarios and acoustic events, and humans critically evaluate, refine, and extend these suggestions. This interactive loop enables the identification of subtle yet important audio-physics cues (e.g., material-dependent impact sounds, fluid dynamics, and environment-induced reverberation) that are often overlooked in existing benchmarks. Through multiple rounds of human verification and consolidation, we obtain a structured pool of candidate phenomena that forms the foundation for subsequent taxonomy design.

\subsection{Taxonomy Construction}

Building upon the surveyed knowledge, we construct the taxonomy of PhyAVBench via a human--LLM co-design process. GPT-5.1 is first prompted to organize the collected phenomena into hierarchical structures, proposing candidate groupings and abstractions. Human experts then review these structures, resolving ambiguities, merging redundant categories, and enforcing consistency with physical principles. This process is repeated iteratively, allowing both the breadth of LLM-generated ideas and the precision of human judgment to be fully leveraged. The resulting taxonomy achieves a balance between coverage and granularity, spanning multiple major dimensions and fine-grained test points. The final taxonomy of PhyAVBench's test points is shown in Table~\ref{tab:test_points}.

\subsection{Prompt Group Design}

After establishing the taxonomy, we design prompt groups through another round of human--LLM collaboration. GPT-5.1 is first used to generate initial prompt candidates for each test point, aiming to produce paired scenarios with controlled variations in physical conditions. 

To further improve the robustness and generalizability of the prompts, we introduce a multi-model cross-validation stage. In this stage, prompts generated by GPT-5.1 are evaluated using GPT-5.1 and Gemini-2.5-pro. We compare the outputs across models to identify inconsistencies, ambiguous phrasings, or cases where the intended physical factors are not consistently preserved. This cross-model agreement process helps filter out unstable or model-specific artifacts and ensures that the prompts are less sensitive to any single model's biases.
The prompt for cross-validation is shown in Fig.~\ref{fig:prompt_validation}.

After cross-validation, human annotators carefully examine the remaining prompts, refining language clarity, eliminating ambiguity, and ensuring that each pair isolates a single audio-physics factor of interest. In addition, discussions between humans and the models are used to identify potential failure modes and edge cases, further improving the discriminative power of the prompts. This iterative refinement process results in high-quality prompt groups that are both diverse and tightly aligned with the intended evaluation goals. The prompt template we use is shown in Fig.~\ref{fig:prompt_generation}.

\end{document}